\documentclass[a4paper,review,11pt,authoryear]{elsarticle}
\usepackage{amsmath,bm,hyperref,url,fullpage,mathtools,booktabs,bbm}
\usepackage{amssymb}
\usepackage{paralist,enumitem,todonotes}
\usepackage{multirow}
\usepackage{threeparttable}
\usepackage{rotating}
\usepackage{float}
\usepackage{graphicx}
\usepackage[ruled,linesnumbered]{algorithm2e}
\usepackage{caption}
\usepackage{subcaption}
\usepackage[a4paper, margin=1in]{geometry}
\usepackage{changes}
\usepackage{enumitem}
\usepackage{algorithm2e}
\usepackage{hyperref}
\hypersetup{
	colorlinks=true,
	linkcolor=blue,
	citecolor=blue,
	urlcolor=blue}
\newtheorem{theorem}{Theorem}  

\newtheorem{lemma}{Lemma}

\newtheorem{remark}{Remark}

\newcommand{\x}{\boldsymbol{x}}
\newcommand{\Nb}{N_b}
\newcommand{\nb}{n_b}
\newcommand{\nt}{n_t}
\newcommand{\D}{\mathcal{D}}
\newcommand{\e}{\epsilon}
\newcommand{\bbeta}{\boldsymbol{\beta}}
\newcommand{\truebeta}{\boldsymbol{\beta}_0}
\newcommand{\oraclebeta}{\hat{\boldsymbol{\beta}}^{\star}_b}
\newcommand{\hatbeta}{\hat{\boldsymbol{\beta}}}
\newcommand{\barbeta}{\bar{\boldsymbol{\beta}}}
\newcommand{\renew}{\tilde{\boldsymbol{\beta}}}
\newcommand{\loss}{\mathcal{L}}
\newcommand{\score}{\mathcal{G}_b}
\newcommand{\grad}{\boldsymbol{g}_b}
\newcommand{\E}{\mathbb{E}}
\newcommand{\renewloss}{\tilde{\loss}}
\newcommand{\Hess}{\nabla^2\loss}
\newcommand{\sumb}{\sum_{t=1}^{b-1}}
\newcommand{\op}{o_p}
\newcommand{\Op}{O_p}
\newcommand{\Sigmaw}{\boldsymbol{\Sigma}_w}

\begin{document}
	\date{}
	\begin{frontmatter}
		\title{Renewable estimation in linear expectile regression models with streaming data sets}
		\author[1]{Wei Cao} 
		\author[1,2]{Shanshan Wang}
		\ead{Corresponding author at (School of Economics and Management, Beihang University, Beijing 100191, China) via sswang@buaa.edu.cn.}
		\author[1]{Xiaoxue Hu}
		
		\address[1]{School of Economics and Management, Beihang
			University, Beijing, China}
		\address[2]{MOE Key Laboratory of Complex System Analysis and Management Decision, Beihang University}
		\begin{abstract}
			Streaming data often exhibit heterogeneity due to heteroscedastic variances or inhomogeneous covariate effects. Online renewable quantile and expectile regression methods provide valuable tools for detecting such heteroscedasticity by combining current data with summary statistics from historical data. However, quantile regression can be computationally demanding because of the non-smooth check function. To address this issue, we propose a novel online renewable method based on expectile regression that efficiently updates estimates using both current observations and historical summaries, thereby reducing storage requirements. Leveraging the smoothness of the expectile loss function, our approach attains superior computational efficiency compared with existing online renewable methods for streaming data exhibiting heteroscedastic variances or inhomogeneous covariate effects. We establish the consistency and asymptotic normality of the proposed estimator under mild regularity conditions, demonstrating that it achieves the same statistical efficiency as oracle estimators based on full individual-level data. Numerical experiments and real-data applications demonstrate that our method performs comparably to the oracle estimator while maintaining high computational efficiency and minimal storage costs. The proposed algorithm has been implemented in the R function \href{https://github.com/Weiccao/ReER}{ReER}.	
		\end{abstract}
		
		\begin{keyword}
			Streaming data; \sep Expectile regression; \sep Renewable estimation
		\end{keyword}
		
	\end{frontmatter}
	
	\section{Introduction}
	\label{sec:introduction}
	
	With rapid technological advances, we have entered the era of large-scale data, characterized by the exponential growth of datasets across diverse fields. Streaming datasets, which arrive continuously over time, have become a key representation of large-scale data \citep{gama2013evaluating}. Streaming data analysis is increasingly essential, with applications in areas such as high-frequency trading, sensor networks, and aerospace. Its dynamic nature imposes higher demands on storage and computation while creating new opportunities and challenges for modeling and analysis \citep{fan2014challenges}. First, storage capacity is limited. As new data continuously arrive, historical raw data are often discarded, making it impractical to store all information. Second, the high velocity of data generation requires real-time processing and rapid analytical updates. Traditional statistical methods, which assume access to complete datasets, are generally unsuitable for this setting. A promising solution is online updating, or renewable estimation, which avoids storing all raw data by updating models using only the current data and summary statistics from the past. Various online updating techniques have been developed, including aggregated estimating equations \citep{lin2011aggregated}, cumulatively updating estimating equations \citep{schifano2016online}, stochastic gradient descent and its variants \citep{chen2020statistical, Zhu02012023}, and renewable estimators \citep{luo2020renewable}.
	
	All of the aforementioned methods and algorithms have demonstrated practical effectiveness. However, they primarily focus on the conditional mean of the response given the predictors or covariates. In many applications, more than just the mean of the conditional distribution is of interest, particularly when covariate effects are inhomogeneous and/or the noise exhibits asymmetric tails. For instance, in the study of the effect of vehicle characteristics on price \citep{fanSequentialQuantileRegression2024a}, one may be more interested in the lower and upper tails rather than the mean of the conditional price distribution given the predictors. To capture heterogeneity in the covariates across different locations of the response distribution, methods such as quantile regression (QR) \citep{koenker1978regression} and asymmetric least squares regression (expectile regression, ER) \citep{newey1987asymmetric} have been widely used.
	
	There has been substantial progress in developing QR methodologies tailored for streaming data settings. For example, \citet{chen2020quantile} proposed an online updating method based on a one-shot algorithm that views streaming data as data blocks partitioned over the time domain. \citet{wang_renewable_2022} developed a renewable QR estimator using a normal approximation of the QR estimator for each batch combined with a maximum likelihood approach. \citet{fanSequentialQuantileRegression2024a} further proposed a sequential Bayesian QR algorithm for streaming data. Recent studies have enhanced online QR methods through smoothing and debiasing techniques in conventional and high-dimensional streaming settings \citep{jiangRenewableQuantileRegression2022,xieStatisticalInferenceSmoothed2024,sunOnlineRenewableSmooth2023,peng2024two}. We refer the reader to \citet{wangselective2025} for a comprehensive overview of recent advances. Despite these advancements, it is computationally challenging due to the non-smooth nature of the check loss. 
	
	Fortunately, ER avoids non-smooth optimization and allows direct estimation of the asymptotic covariance matrix without requiring nonparametric density estimation, leading to greater computational efficiency. While QR focuses on tail probabilities, ER captures both the likelihood and magnitude of extreme values, making it particularly suitable for analyzing high-impact risks. For example, \citet{taylor2008estimating} introduced expectile-based Value-at-Risk to better assess rare, high-loss events, and \citet{kuan_assessing_2009} proposed a conditional autoregressive expectile model. More recent studies, such as \citet{mohammedi2021consistency} and \citet{sahamkhadam2021dynamic}, have further advanced expectile-based risk measures. Consequently, ER has been widely applied in areas such as risk measurement \citep{kim2016nonlinear, daouia_estimation_2018, xu_prediction_2022}, exchange rate volatility \citep{xie2014varying}, stock index modeling \citep{jiang2022single}, and high-frequency financial data analysis \citep{gerlach2022bayesian}.
	
	Thanks to its advantages and broad applicability, expectile regression (ER) for large-scale data has recently become an active research area in modern statistics. To address computational challenges, researchers have developed a variety of distributed and subsampling-based methods for ER models. We refer readers to \citet{song2021linear,hu2022distributedexpectile,pan2021distributed,pan2025distributed} for distributed ER, and to \citet{chen2024estimation,li2024poisson,li2025efficient} for subsampling-based ER. However, ER for streaming data remains largely unexplored, and the existing above methods cannot be directly applied to streaming settings.
	
	Inspired by the one-shot approach, \citet{pan2024renewable} proposed an online updating strategy for ER that constructs a joint likelihood function for each data batch and optimizes it using the Alternating Direction Method of Multipliers (ADMM). To the best of our knowledge, it is the latest literature working on this topic. However, its renewable estimator is based on a one-shot strategy, enabling ER estimation for streaming data but fails to leverage cross-batch information, resulting in inefficient use of available data. Moreover, their asymptotic theory requires that each block size be larger than $\sqrt{N_b}$ and diverge to infinity, where $N_b$ denotes the total sample size. This assumption may not hold in real-world scenarios where data arrive in small or unevenly sized batches. We also demonstrate this phenomenon in our simulation studies.
	
	In this study, we propose a \underline{Re}newable \underline{E}xpectile \underline{R}egression estimation strategy for the streaming datasets, hereafter referred to as \textbf{ReER}. Unlike the method of \citet{pan2024renewable}, our approach builds on the online updating framework introduced by \citet{luo2020renewable}, reconstructing the renewable ER loss function via Taylor expansion and deriving an approximate iterative solution. The main contributions and advantages of the proposed method can be summarized as follows. First, from an estimation perspective, \textbf{ReER} employs Taylor expansion to refine historical parameter estimates using newly arrived data, enabling efficient information integration across batches and overcoming the limitations of one-shot estimation.
	Second, from a computational perspective, \textbf{ReER} requires only the current data batch and a set of sufficient statistics from previous batches, substantially reducing memory and computational costs.
	Finally, from a theoretical perspective, the proposed estimator achieves asymptotic efficiency comparable to that of the full-sample estimator under mild conditions. Specifically, it only requires a consistent initial ER estimator and a sufficiently large first-batch sample size, without imposing additional distributional or structural assumptions.
	Overall, the \textbf{ReER} method is computationally efficient in handling the challenges of data storage and processing while maintaining asymptotic equivalence to the oracle ER estimator.

	The remainder of this article is organized as follows. Section~\ref{sec:method} introduces the renewable method for estimating ER parameters and establishes its large-sample properties in Section \ref{sec:properties}. Section~\ref{sec:experiments} evaluates the performance of the proposed method through numerical simulations. Two real-data applications are presented in Section~\ref{sec:realdata}. Section~\ref{sec:discussion} provides a brief conclusion and discusses potential future research directions for online updating in ER. All technical details are provided in the Appendix.
	
	\section{Methodology}\label{sec:method}
	
	We first introduce the linear ER model and relevant notations in Section~\ref{sub:2-1}, followed by the derivation of the proposed \textbf{ReER} estimator and its corresponding algorithm in Section~\ref{sub:2-2}. Section~\ref{sub:2-3} presents a comparison between the proposed \textbf{ReER} method and those of \citet{pan2024renewable} and \citet{song2021linear}, and Section~\ref{sub:2-4} discusses its computational complexity.
	
	\subsection{Model and Notations}\label{sub:2-1}
	
	For $\tau\in(0,1)$, let $e_\tau(Y|X=\x)$ denote the conditional expectile of $Y$ given $X=\x$ at $\tau$th expectile, which can be obtained by minimizing the asymmetric squared loss function \citep{newey1987asymmetric}:
	\begin{align}\label{quantile}
		e_\tau(Y|X=\x) = \mathop{\arg\min}\limits_{\theta} E\Big(\rho_\tau(Y-\theta)|X=\x\Big),
	\end{align}
	where $\rho_\tau(u)=\frac{1}{2}\cdot u^2\cdot|\tau-I(u<0)|$, and $I(u<0)$ is an indicator function. Assume that $\D_{(b)}=\{\D_t: t=1,2,\cdots,b\}$ represents the streaming dataset up to batch $b$, where the total sample size is $\Nb=\sum_{t=1}^b \nt$. For each batch $\D_t$, it consists of observations $\{(\x_{ti},y_{ti})\}_{i=1}^{\nt}$, where $\x_{ti}=(x_{ti,1},\cdots,x_{ti,p})^\top$ is a $p \times 1$  covariate vector, and $\nt$ is the sample size of batch $t$. The data follows the linear expectile regression model:
	\begin{align}
		\label{eq:expectile}
		e_\tau(y_{ti}|\x_{ti})=\x_{ti}^\top \bbeta(\tau), \quad i=1,\cdots, t.
	\end{align}
	where $\bbeta (\tau)$ is the unknown expectile regression parameter. To simplify notation, we omit  $\tau$  where there is no ambiguity. Additionally, we use: ``$\tilde{\ \ }$'' to indicate online estimators, e.g.,  $\renew_{b}$  represents the renewable estimator based on data up to batch  $b$. ``$\hat{\ \ }$''  to indicate static estimators, e.g.,  $\hatbeta_t$  is the estimator computed only from batch $\D_t$. 
	
	\subsection{ReER: Renewable estimation for Expectile Regression}\label{sub:2-2}
	
	First, the conventional oracle expectile regression estimator of $\bbeta$, i.e., $\oraclebeta$, can be obtained by minimizing the following global loss function: 
	\begin{align}\label{eq:oracle-est}
		\begin{aligned}
			\oraclebeta=\mathop{\arg\min}\limits_{\bbeta} \ \loss_{\Nb}(\bbeta),\ \  & \loss_{\Nb}(\bbeta)=
			\frac{1}{\Nb}\sum_{t=1}^{b} \nt \cdot \loss_{\nt}(\bbeta),
		\end{aligned}
	\end{align}
	where $\loss_{\nt}(\bbeta)$ is the loss function only obtained based on samples from batch $\D_t$, i.e.,
	\begin{align}\label{eq:loss-oracle-t}
		\begin{aligned}
			\loss_{\nt}(\bbeta)&=
			\frac{1}{\nt}\sum_{i=1}^{\nt} \rho_{\tau}(y_{ti}-\x_{ti}^{\top}\bbeta).
		\end{aligned}
	\end{align}
	The Eq.~\eqref{eq:oracle-est} can be solved using an iteratively reweighted least squares (IRLS) method, and the solution of the oracle ER estimator $\oraclebeta$ is given by:
	\begin{align}\label{eq:oracle-sol}
		\oraclebeta=\big[W_{\Nb}(\oraclebeta)\big]^{-1}U_{\Nb}(\oraclebeta),
	\end{align}
	where $W_{N_b}(\oraclebeta)$ and $U_{N_b}(\oraclebeta)$ are: 
	\begin{align}\label{eq:IWLS-1}
		W_{N_b}(\oraclebeta)=\sum_{t=1}^{b}\sum_{i=1}^{\nt} |\tau-I\big(y_{ti}<\x_{ti}^{\top}\oraclebeta\big)|\x_{ti}\x_{ti}^{\top},\ \ 
		U_{N_b}(\oraclebeta)=\sum_{t=1}^{b}\sum_{i=1}^{\nt} |\tau-I\big(y_{ti}<\x_{ti}^{\top}\oraclebeta\big)|\x_{ti}y_{ti}.  
	\end{align}
	
	If we could aggregate all information from the all $b$ data batches at once, $\oraclebeta$ could be obtained by solving Eqs.~\eqref{eq:oracle-sol} and \eqref{eq:IWLS-1}. However, due to the large cumulative sample size and memory constraints, this is often infeasible in practice. To address this, next, we propose a renewable ER estimator that relies only on the current data batch and a few summary statistics.
	
	Let $\loss_{\Nb}(\bbeta)$ and $\loss_{\nb}(\bbeta)$ represent the overall loss functions based on the data up to the $b$-th batch (i.e., $\D_{(b)}$) and the $b$-th batch of data (i.e., $\D_b$), respectively. We begin with a simple scenario of two batches of data $\mathcal{D}_1$ and $\mathcal{D}_2$, where $\mathcal{D}_2$ arrives after $\mathcal{D}_1$. First we estimate the initial ER estimate $\hatbeta_1$ (or $\hatbeta_1^*$, $\renew_1$) with $\D_1$. Next we update the initial ER estimate to a renewed ER estimate $\renew_2$ without using any subject level data but only some summary statistics from $\D_1$. Following \citet{luo2020renewable}, the oracle ER estimator $\hat{\boldsymbol{\beta}}_2^*$ is obtained by minimizing the following aggregated loss function:
	\begin{align}\label{loss-b2}
		\loss_{N_2}(\bbeta)&=\frac{1}{N_2}\Big(n_1\loss_{n_1}(\bbeta)+n_2\loss_{n_2}(\bbeta)\Big).
	\end{align}
	The minimizer of \eqref{loss-b2} is equivalent to the minimizer of the following loss function:
	\begin{align}\label{loss-b2-1}
		\loss_{N_2}(\bbeta)&=\frac{1}{N_2}\left[n_1\left(\loss_{n_1}(\bbeta)-\loss_{n_1}(\tilde{\boldsymbol{\beta}}_1)\right)+n_2\loss_{n_2}(\bbeta)\right].
	\end{align}
	Assume that conditions (C1)-(C4) hold (see details in Section~\ref{sub:2-3}), the loss function (\ref{loss-b2-1}) is asymptotically equivalent to: 
	\begin{align}\label{loss-b2-2}
		\frac{n_1}{2N_2}(\bbeta-\renew_1)^\top\Hess_{n_1}(\renew_{1})(\bbeta-\renew_1)+\frac{n_2}{N_2}\loss_{n_2}(\bbeta),
	\end{align}
	where $\Hess_{n_1}(\renew_{1})$ is the Hessian matrix of the loss function of data batch $\D_1$. Thus, when data batch $\D_2$ arrives, we can obtain the renewable estimator $\renew_2$ by minimizing (\ref{loss-b2-2}), which only depends only on current data batch $\D_2$, summary statistics $n_1, \renew_1$ and $\Hess_{n_1}(\renew_{1})$ of batch data $\D_1$. 
	
	Similar, when the $b$-th batch arrives, the overall loss functions $\loss_{N_b}(\bbeta)$ can be represented by an aggregated form with historical data part and the newly arrived data part:
	\begin{align}\label{eq:loss-decompose}
		\loss_{\Nb}(\bbeta)=\frac{1}{\Nb}\Big(N_{b-1}\loss_{N_{b-1}}(\bbeta)+\nb\loss_{\nb}(\bbeta)\Big).
	\end{align}
	Eq.~\eqref{eq:loss-decompose} separates the overall loss into components attributable to historical data and the newly arriving data, enabling an efficient update mechanism for renewable estimation. Performing a Taylor expansion around the online estimator $\renew_{b-1}$ on the historical term of \eqref{eq:loss-decompose}, the loss function $\loss_{\Nb}(\bbeta)$ can be approximately represented as:
	\begin{align} \label{eq:online-taylor}
		\loss_{N_b}(\bbeta)\approx \renewloss_{\Nb}(\bbeta)=\frac{1}{\Nb}\Bigg\{\frac{1}{2}(\bbeta-\renew_{b-1})^\top\Bigg[\sumb\nt \Hess_{\nt}(\renew_{t})\Bigg](\bbeta-\renew_{b-1})+n_b\loss_{\nb}(\bbeta)\Bigg\},
	\end{align}
	where $\Hess_{\nt}(\renew_{t})$ is the Hessian matrix of the loss function of data batch $t$. 
	
	As mentioned earlier, compared to QR, the ER model employs an asymmetric squared loss, which allows it to be directly solved via differentiation. Then minimizing (\ref{eq:online-taylor}) yields the following updated estimator:
	\begin{align} \label{eq:online-sol}
		\renew_b = \left[ \sumb W_{\nt}(\renew_t) + W_{\nb}(\renew_b) \right]^{-1} \left[ \sumb W_{\nt}(\renew_t) \renew_{b-1} + U_{\nb}(\renew_b) \right],
	\end{align}
	where $W_{\nt}(\renew_t)$ and  $U_{\nt}(\renew_t)$ are defined as:
	$$W_{\nt}(\renew_t)=\sum_{i=1}^{\nt} |\tau-I(y_{ti}<\x_{ti}^{\top}\renew_t)|\x_{ti}\x_{ti}^\top,\ \ U_{\nt}(\renew_t)=\sum_{i=1}^{\nt} |\tau-I(y_{ti}<\x_{ti}^{\top}\renew_t)|\x_{ti}y_{ti}.$$
	
	Note that the right-hand side of Eq.~\eqref{eq:online-sol} still involves the unknown parameter $\renew_b$, and thus does not yield a closed-form solution for $\renew_b$. To reduce computational burden, we replace $\renew_b$ with a consistent initial estimator on the right-hand side of Eq.~\eqref{eq:online-sol}. Specifically, we approximate $\renew_b$ using the renewable estimator from the previous step, $\renew_{b-1}$, which was obtained from the preceding data batch $\D_{(b-1)}$. Accordingly, $W_{\nb}(\renew_{b})$ and $U_{\nb}(\renew_{b})$ are approximated by $W_{\nb}(\renew_{b-1})$ and $U_{\nb}(\renew_{b-1})$, respectively. Thus, in practice, the final estimator is given by:
	\begin{align}
		\label{eq:renewable}
		\renew_b&=\left[ \sumb W_{\nt}(\renew_t) + W_{\nb}(\renew_{b-1}) \right]^{-1} \left[\sumb W_{\nt}(\renew_t) \renew_{b-1} + U_{\nb}(\renew_{b-1}) \right].
	\end{align}
	Thus the renewable estimator $\renew_b$ can be computed using only the aggregated summary statistics from previous data batches, specifically $\left\{\mathbb{H}_{N_{b-1}}(\renew), \renew_{b-1} \right\}$, along with the statistics derived from the newly arrived data batch $\left\{W_{\nb}(\renew_{b-1}), U_{\nb}(\renew_{b-1})\right\}$, where $\mathbb{H}_{N_{b-1}}(\renew) = \sumb W_{\nt}(\renew_t)$.
	
	\begin{figure}[ht]
		\centering
		\includegraphics[width=1\linewidth] {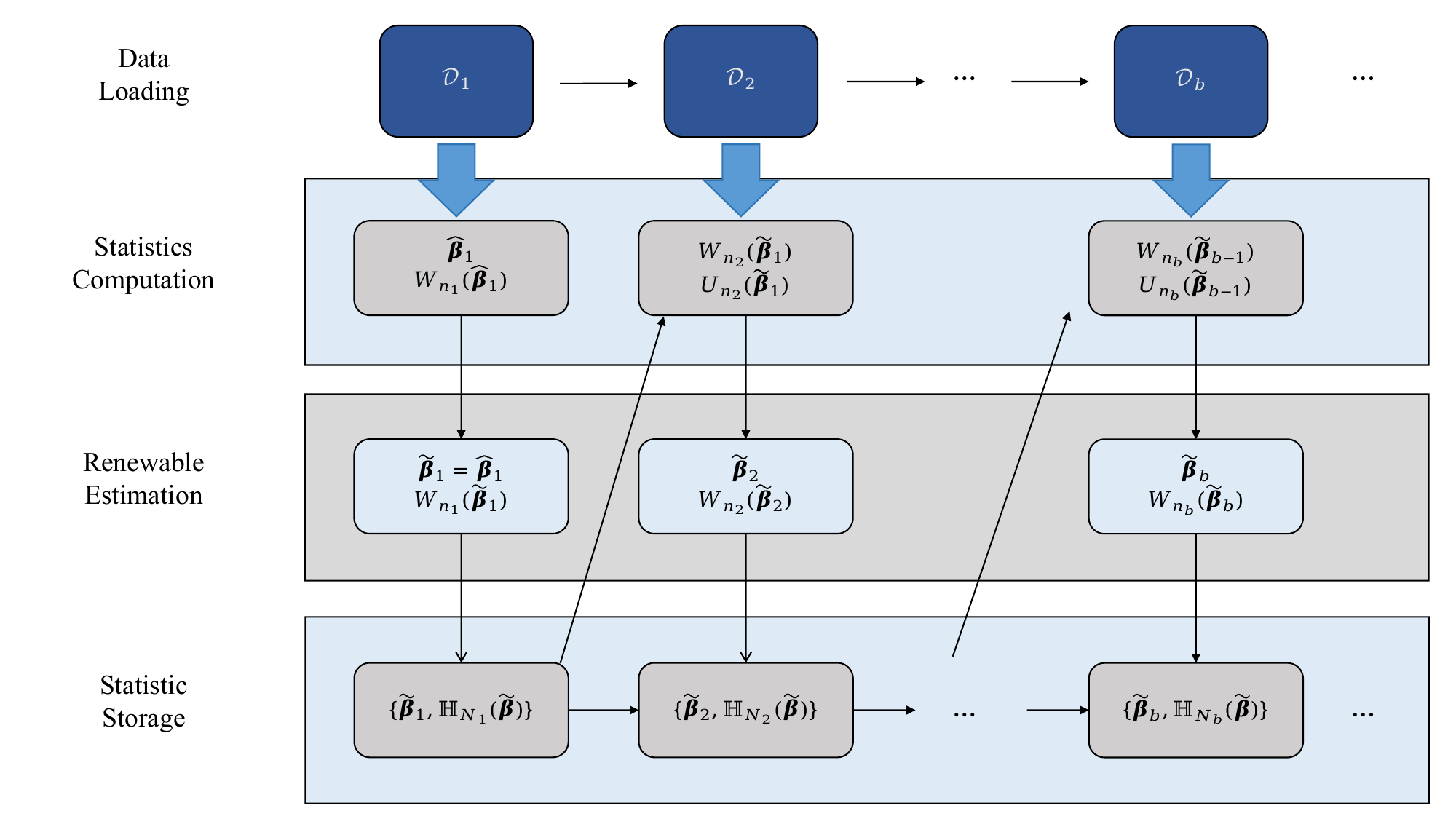}
		\caption{Illustration of the proposed online renewable ReER.}
		\label{fig:frame}
	\end{figure}
	
	A clear overview of the proposed renewable estimation framework is illustrated in Figure~\ref{fig:frame}. Specifically, when the $b$-th data batch arrives, the \textbf{ReER} algorithm requires only the previous renewable estimator $\renew_{b-1}$, and the current data batch $\D_b = \{\x_{bi}, y_{bi}\}_{i=1}^{\nb}$, to compute the necessary summary statistics. During this step, the updated estimator is obtained by combining the statistics of the current batch with the cumulative Hessian matrix $\mathbb{H}_{N_{b-1}}(\renew)$ from the previous batches. Specifically, upon the arrival of the $b$-th data batch, the only summary information that needs to be retained is $\mathcal{S}_b = \{\mathbb{H}_{N_b}(\renew), \renew_b\}$. This implies that the algorithm stores only a $p \times p$ matrix and a $p \times 1$ vector, as opposed to the full $\Nb \times p$ data matrix, which can be prohibitively large. This compact representation becomes especially advantageous as the number of data batches grows while the parameter dimension $p$ remains moderate. 
	In summary, the proposed \textbf{ReER} method updates model parameters using only the current batch and summary information from past data, eliminating the need to store the entire historical dataset. This makes it computationally efficient and well-suited for large-scale streaming data analysis. Finally, 
	the main steps of the proposed \textbf{ReER} procedure are further illustrated in Algorithm~\ref{alg:ReER-alg}.
	
	\begin{algorithm}[ht]
		\linespread{0.9}\selectfont
		\caption{ReER: Renewable estimation in expectile regression}
		\label{alg:ReER-alg}
		\SetKwData{Left}{left}\SetKwData{This}{this}\SetKwData{Up}{up}
		\SetKwFunction{Union}{Union}\SetKwFunction{FindCompress}{FindCompress}
		\SetKwInOut{Input}{Input}\SetKwInOut{Output}{Output}
		\Input{Observed data of data batch $\D_t:\{\x_{ti},y_{ti}\}_{i=1}^{\nt}$, expectile levels $\tau$.}
		\Output{The $\tau$th renewable expectile estimator: 
			$\renew_t(\tau)$}
		
		\textbf{Initialize:}
		Initial the renew ER estimator with first batch $\D_1$ by optimize Eq.~\eqref{eq:loss-oracle-t}:
		$$\renew_1=\hatbeta_1=\mathop{\arg\min}\limits_{\boldsymbol{\beta}} \ \loss_{n_1}(\bbeta)=\frac{1}{n_1}\sum_{i=1}^{n_1} \rho_{\tau}\big(y_{i1}-\x_{i1}^\top\bbeta\big),$$\\
		
		Calculate and storage $\mathcal{S}_1$: 
		$$\mathcal{S}_1=\left\{W_{n_1}(\renew_1), \renew_1\right\}, W_{n_1}(\renew_1)=\sum_{i=1}^{\nt} |\tau-I(y_{1i}<\x_{1i}^{\top}\renew_1)|\boldsymbol{x}_{1i}\x_{1i}^\top.$$
		
		\For{$t=2$ \KwTo $b$}{
			Calculate the renewable estimator  $\renew_t$ by Eq.~\eqref{eq:renewable}\\
			Update the historical statistic (cumulative Hessian matrix $\mathbb{H}_{N_t}$) by:
			$$\mathbb{H}_{N_t}(\renew)=\mathbb{H}_{N_{t-1}}(\renew)+W_{t}(\renew).$$\\
			
			Storage the summary information up to data batch $t$:
			$$\mathcal{S}_b=\Big\{\mathbb{H}_{\Nb}(\renew), \renew_b\Big\}$$}
	\end{algorithm}
	
	\subsection{Some remarks}\label{sub:2-3}
	
	Here we discuss the key differences between the proposed ReER method and those of \citet{pan2024renewable} and \citet{song2021linear}, respectively.
	
	Our proposed \textbf{ReER} method differs from the online ER method proposed by \citet{pan2024renewable} in several key aspects. First, the two approaches diverge in the estimation of local static estimators. In the initialization step of \textbf{ReER}, the estimator $\renew_1$ is obtained by minimizing the expectile loss over the first arriving data batch $\D_1$. This optimization problem is solved using IRLS algorithm, which is implemented in \textsc{R} via the \texttt{expectreg} package. In contrast, the method of \citet{pan2024renewable} employs ADMM for optimization. To ensure comparability between the two methods, we adopt the IRLS algorithm for both approaches in the remainder of this paper. Second, the structure of the renewable estimator also differs between the two methods. The renewable estimator proposed by \citet{pan2024renewable} is defined as:
	\begin{align}
		\label{eq:PAER}
		\renew_b^{(\text{pan})}=\left[\sum_{t=1}^b\omega_t\frac{\boldsymbol{X_t}^\top\boldsymbol{X}_t}{\nt}\right]^{-1}\left[\sum_{t=1}^b\omega_t\frac{\boldsymbol{X_t}^\top\boldsymbol{X}_t}{
			\nt}\hatbeta_t\right],
	\end{align}
	where $\omega_t = n_t / N_t$. This estimator essentially performs a weighted aggregation of local estimators from each data batch, with weights determined by both covariance matrices and the sample size of each batch. As mentioned earlier, ER can be viewed as an extension of OLS regression that incorporates an asymmetric loss function depending on the expectile level $\tau$. In Eq.~\eqref{eq:renewable}, the statistics $W_{n_t}(\renew_t)$ and $U_{n_t}(\renew_t)$ inherently reflect this asymmetry, as they are computed under the expectile loss framework. Therefore, the proposed method effectively captures cross-batch information, including the heterogeneity among batches, making it particularly well-suited for scenarios with small or limited batch sizes.
	
	\citet{song2021linear} raised a ER algorithm for massive data under divide an conquer (DC) scenario, which the estimator can be derived by:  
	\begin{align}
		\renew_t(\tau)=\left[\sum_{t=1}^{b}\widehat{Q}_t^{-1}(\hatbeta_t)\right]^{-1}\left[\sum_{t=1}^{b}\widehat{Q}_t^{-1}(\hatbeta_t)\hatbeta_t\right],
		\label{eq:dcer}
	\end{align}
	where $\widehat{Q}_t(\hatbeta_t)$ is the plug-in estimator of the asymptotic covariance matrix (for more detail see \citet{song2021linear}). This DC-based method can be naturally extended to streaming scenarios, as the structure of Eq.~\eqref{eq:dcer} is a sum-linear form over individual data batches. Compared to the method of \citet{pan2024renewable}, the approach of \citet{song2021linear} also incorporates expectile loss information. However, there are still important differences between their method and the proposed \textbf{ReER}. The renewable estimator proposed by \citet{song2021linear} is essentially a variant of the DC-based estimator and it is still a weighted average of local static estimates. In contrast, our proposed renewable estimator \textbf{ReER} is based on an update-oriented framework. Specifically, it incrementally updates the parameter estimates using the newly arrived data batch along with cumulative summary statistics from historical data. This updating mechanism enables \textbf{ReER} to more effectively capture cross-batch information and ensures robustness even when the incoming data batch significantly differs from previous ones or when batch sizes are small. Consequently, the proposed estimator tends to be more robust in heterogeneous or low-sample-size scenarios.
	
	\subsection{Computation complexity}\label{sub:2-4}
	
	In this section, we take a preliminary analysis of the computational complexity of Oracle ER and the proposed \textbf{ReER}. For simplicity, we assume the samples size of each data batch is same, which $n_t=n=\Nb/b$ for $t=1,\cdots,b$.
	
	For the Oracle and local estimations, we apply the \textsf{expectreg.ls} function from the \texttt{expectreg} package in R, which employs the IRLS method. Assuming the number of IRLS iterations is $T$, the computational complexity of computing the oracle estimator $\oraclebeta$ and the local (initial) estimator $\hatbeta_1$ is approximately $O\left(T(\Nb \cdot p^2 + p^3)\right)$ and $O\left(T(n \cdot p^2 + p^3)\right)$, respectively.
	
	For the \textbf{ReER} method, computing $W_{\nt}(\renew_t)$ and $U_{\nt}(\renew_t)$ requires $O(np^2)$ and $O(np)$, respectively. The total computational cost for updating $\tilde{\boldsymbol{\beta}}_b$ is $O(np^2 + np + p^2 + p^3)$, where the $O(p^2)$ term arises from the computation of $\sumb W_{\nt}(\renew_t)\renew_{b-1}$, and the $O(p^3)$ term comes from inverting $\left[ \sumb W_{\nt}(\renew_t) + W_{\nb}(\renew_{b-1}) \right]$. Considering updates over $b$ data batches, the overall computational complexity of the \textbf{ReER} method is:
	$O\left( T(np^2 + p^3) + b(np^2 + np + p^2 + p^3) \right)$.
	
	In practice, the total sample size $\Nb$ can exceed one million. Meanwhile, the IRLS algorithm typically converges within a small number of iterations, often fewer than 10, implying that $T \ll N$ in most applications. Furthermore, we focus on a low-dimensional setting where the number of covariates $p$ satisfies $p \ll N$. The dominant term in the complexity of Oracle estimator is $O(T\Nb\cdot p^2)$, while \textbf{ReER}, it is $O(Tn\cdot p^2+\Nb \cdot p^2)$. Due to the fact that $n \ll \Nb$, \textbf{ReER} greatly saves the computational cost compared to Oracle estimator.

	\section{Large Sample Properties}\label{sec:properties}
	
	In this section, we present the large sample properties for the \textbf{ReER} method. To establish the theoretical results, we impose some notations. Define $\lambda_{\min}(\boldsymbol{A})$ and $\lambda_{\max}(\boldsymbol{A})$ as the smallest and largest eigenvalues of a square matrix $\boldsymbol{A}$. 
	For a vector, $\|\cdot\|$ denotes its $l_2$-norm, and for a matrix, $\|\cdot\|$ denotes its spectral norm. And let $c_l$ be positive constant, which $l=1,2,3$ and $4$. The following regular conditions are required:
	
	\begin{enumerate} [label=(C\arabic*)] 
		\item The unknown parameter vector $\boldsymbol{\beta}_0$ lies in the interior of a compact and convex parameter space $\mathcal{H}$, which is a subset of $\mathbb{R}^{p}$.
		
		\item  Let $\e_{ti}=y_{ti}-\x_{ti}^\top\truebeta$ be the error tern of ER model. Given $X=\x$, $\e_{ti}$ are i.i.d. distributed and satisfy $e_\tau(\e_{ti}|X=\x)=0$ and $\E\Big[\e_{ti}^2|X=\x\Big] \leq c_1 <\infty$. 
		
		\item Define $\boldsymbol{\Sigma}=\E[\x_{ti}\x_{ti}^\top]$, and there exists a positive constant $c_2$ such that $0 <\lambda_{\min}(\boldsymbol{\Sigma})<\lambda_{\max}(\boldsymbol{\Sigma})\leq c_2< \infty$.
	\end{enumerate}
	
	Conditions (C1)–(C3) are standard assumptions for ER estimator, as discussed in \citet{newey1987asymmetric}. Under these conditions the oracle estimator $\oraclebeta$ satisfy the following asymptotic distribution when $\Nb\rightarrow\infty$:
	$$\sqrt{\Nb}\big(\oraclebeta-\truebeta\big) \xrightarrow{d}\big(\textbf{0},\Sigmaw^{-1}\boldsymbol{\Omega}\Sigmaw^{-1} \big),$$
	where $\boldsymbol{\Omega}=\E\Big[|\tau-1(\e_{ti}<0)|^2\cdot\e_{ti}^2\cdot\x_{ti}\x_{ti}^\top\Big]$ and $\Sigmaw=\E\Big[|\tau-1(\e_{ti}<0)|\x_{ti}\x_{ti}^\top\Big]$. 
	
	Since the renewable estimator $\renew_b$ is obtained by optimizing Eq.~\eqref{eq:online-taylor}, we impose the following regularity condition to facilitate the investigation of its theoretical properties:
	
	\begin{enumerate} [label=(C\arabic*),start = 4] 
		\item  Given $X=\x$, the conditional density function $f_{\epsilon_{ti}}(\cdot)$ of error term $\epsilon_{ti}$ is bounded, satisfy $ 0 < f_{\epsilon_{ti}}(\cdot) \leq c_3 < \infty$. 
		\item The sample size of the first data batch $n_1 \rightarrow \infty$, and for $t=1,2,\cdots$ and $i=1,2,\cdots, n_t$, there exists a positive constant $c_4$ such that $\|\x_{ti}\|<c_4$
	\end{enumerate}
	
	This condition ensures the consistency of the initial estimator for the first batch, $\renew_1$. When $n_1 \rightarrow \infty$, under conditions (C1)–(C3), we have $\hatbeta_1=\renew_1\stackrel{p}{\rightarrow} \boldsymbol{\beta}_0$. Moreover, since $\nabla\loss_{n_1}(\renew_1) = \mathbf{0}$, the approximation in Eq.~\eqref{eq:online-taylor} holds. The consistency and asymptotic normality of the proposed \textbf{ReER} estimator are established in Theorems~\ref{T1} and ~\ref{T2}. In addition, the asymptotically equivalent of the proposed renewable estimator $\renew_b$ and the oracle estimator $\oraclebeta$ are given in Theorem~\ref{T3}. The proof details are given in \ref{app:proof}.
	
	\begin{theorem}\label{T1}
		{\color{blue}(Consistency of $\tilde{\boldsymbol{\beta}}_b$)}. Assume that conditions (C1)-(C5) hold, the renewable estimator $\renew_b$ is consistent with the true parameter, that is:
		$$\renew_b \xrightarrow{p} \truebeta.$$
	\end{theorem}
	
	\begin{theorem}\label{T2}
		{\color{blue}(Asymptotic Normality of $\renew_b$)}. Assume that conditions (C1)-(C5) hold, $\renew_b$ satisfies the following asymptotic normal distribution:
		$$\sqrt{N_b}(\renew_b-\bbeta_0) \xrightarrow{d}\big(\textbf{0},\Sigmaw^{-1}\boldsymbol{\Omega}\Sigmaw^{-1}\big).$$
	\end{theorem}
	
	\begin{theorem}\label{T3}
		{\color{blue}(Asymptotic equivalency of $\renew_b$)}. Under conditions (C1)-(C5), the $l_2$-norm difference between the oracle estimator and renewable estimator vanishes at the rate of $\Nb^{-1}$, i.e.,
		$$\|\renew_b-\oraclebeta\| = O_p(1/\Nb).$$
	\end{theorem}
	\begin{remark}
		In streaming data analysis, it is standard and commonly assumed that the number of data batches satisfies $b \rightarrow \infty$ in real applications \citep{schifano2016online}. Many existing studies further assume that $b = O(N_b^{\kappa})$ for some constant $0 < \kappa < 1$, which implicitly requires  the batch sizes $\nt \rightarrow \infty$ for all $t = 1, 2, \dots$ \citep{song2021linear, pan2024renewable}. However, such assumptions may not hold in many real-world scenarios where data arrive continuously over time with fixed or limited batch sizes. In contrast, our proposed method only requires that the first batch size satisfies $n_1 \rightarrow \infty$, making it more flexible and well-suited for practical applications involving perpetual data streams. This advantage is also supported by both simulation studies and real data applications.    
	\end{remark}
	
	\section{Simulation Studies}
	\label{sec:experiments}
	
	We first describe the model setup and evaluation criteria in Section~\ref{sub:3-1}, and then examine the effects of batch size and the number of batches on finite-sample performance in Section~\ref{sub:3-2}. Finally, Section~\ref{sub:3-3} evaluates the computational efficiency of the proposed method.
	
	\subsection{Model setup and evaluation criteria}\label{sub:3-1}
	
	Assume that data are generated from the following model:
	\begin{align}\label{eq:model}
		y_{ti} = \x_{ti}^{\top}\bbeta^{\star}+(\x_{ti}^{\top}\boldsymbol{\gamma}) \e_{ti}, \quad t=1,\cdots, b,
	\end{align}
	where $\x_{ti}=(1,x_{ti,1},x_{ti,2})^\top$ and $x_{ti,1},x_{ti,2}$ follow a uniform distribution $\text{Unif}(0,1)$. According to Eq.~\eqref{eq:model}, the true expected expectile regression coefficient is $\truebeta(\tau)=\bbeta^{\star}+e_{\tau}(\boldsymbol{\e})\cdot \boldsymbol{\gamma}$, where $e_{\tau}(\e)$ denotes the $\tau$th expectile of error terms. Note that when $\x_{ti}^{\top}\boldsymbol{\gamma}$ is constant, model \eqref{eq:model} has homogeneous variance; otherwise, the errors are heteroscedastic. For the random error $\e_{ti}$, we consider the following two cases: 1) $\e_{ti}$ follows the standard normal distribution, i.e., $\e_{ti}\sim N(0,1)$ and 2) $\e_{ti}$ follows the student $t$ distribution with $3$ degree of freedom, i.e., $\e_{ti}\sim t(3)$. We set the true $\bbeta^\star=(2,1,2)^\top$ with $\boldsymbol{\gamma}=(1,0,0)^\top$ for the homogeneous model and $\boldsymbol{\gamma}=(1,0.25,0)^\top$ for the heterogeneous model. Thus, we consider the following four cases based on the homogeneity or heterogeneity of the model and the distribution of the error term:
	
	\begin{itemize}[label={},leftmargin=0pt, itemindent=0pt, labelsep=0pt]
		\item \textbf{Case 1}: Homogeneous model with error term $\boldsymbol{\e} \sim N(0,1)$;
		\item \textbf{Case 2}: Homogeneous model with error term $\boldsymbol{\e} \sim t(3)$;
		\item \textbf{Case 3}: Heterogeneous model with error term $\boldsymbol{\e} \sim N(0,1)$;
		\item \textbf{Case 4}: Heterogeneous model with error term $\boldsymbol{\e} \sim t(3)$.
	\end{itemize}
	
	For each case, we consider these two scenarios:
	\begin{itemize}[label={}, leftmargin=0pt, itemindent=0pt, labelsep=0pt]
		\item \textbf{Scenario S1}: We fix $N_b=100,000$ and vary the batch size $n_k$ to evaluate the impact of batch size on performance.
		\item \textbf{Scenario S2}: We fix $n_k=300$ and vary the number of batches $K$ to evaluate the number of batches on performance.
	\end{itemize}
	
	For a comprehensive comparison, we evaluate our proposed method against three benchmark estimators from the existing literature:
	(i) \textbf{Oracle}: estimation using the entire dataset as a gold standard; (ii) \textbf{DCER}: a divide-and-conquer-based estimator for expectile regression proposed by \citet{song2021linear}; and (iii) \textbf{PAER}: a renewable estimation approach for expectile regression introduced by \citet{pan2024renewable}. And we conduct simulation $200$ times. The main performance metrics are: (1) the empirical bias (\textbf{BIAS}), defined as $\textbf{BIAS}(\beta_j)=\frac{1}{200}\sum_{s=1}^{200}(\tilde{\beta}^{(s)}_j-\beta_j)$, (2) the mean square error (\textbf{MSE}), i.e., $\textbf{MSE}(\beta_j)=\frac{1}{200}\sum_{s=1}^{200}\big(\tilde{\beta}^{(s)}_j-\beta_j\big)^2$ and computational time on average (\textbf{Time}), which is measured after the final data batch is processed. Here, $\tilde{\beta}^{(s)}_j$ denotes the renewable estimator of $\beta_j$ ($j=0,1,2$) in the $s$th repetition, with $s = 1, 2, \ldots, 200$.

	Throughout this paper, the proposed renewable estimation method is referred to as \textbf{ReER}. All experiments are conducted in R, an R package to implement our proposed method can be obtained at the following link: \href{https://github.com/Weiccao/ReER}{https://github.com/Weiccao/ReER}. Due to space limitations, we report results only for the case where $\tau = 0.25$.  Additional results corresponding to $\tau=0.5$ and $\tau = 0.75$ expectile levels are presented in \ref{app:sim}.
	
	\subsection{Performance evaluation}\label{sub:3-2}
	
	\subsubsection{Evaluation of the batch size on performance}\label{S1}
	
	Scenario \textbf{S1} is designed to evaluate the impact of the data batch size $n_k$ on the statistical performance of the proposed renewable estimators. Here we fix the total sample size at $N_k = 100,000$ and vary the batch size $n_k$, where $n_k$  $\in \{200,1000,2000,5000,10000\}$. The simulation results are presented in Tables \ref{tab:fix-homo}-\ref{tab:fix-hetero} for Cases 1-4, respectively.
	
\begin{figure}[htbp]
		\centering
		\includegraphics[width=\linewidth,]{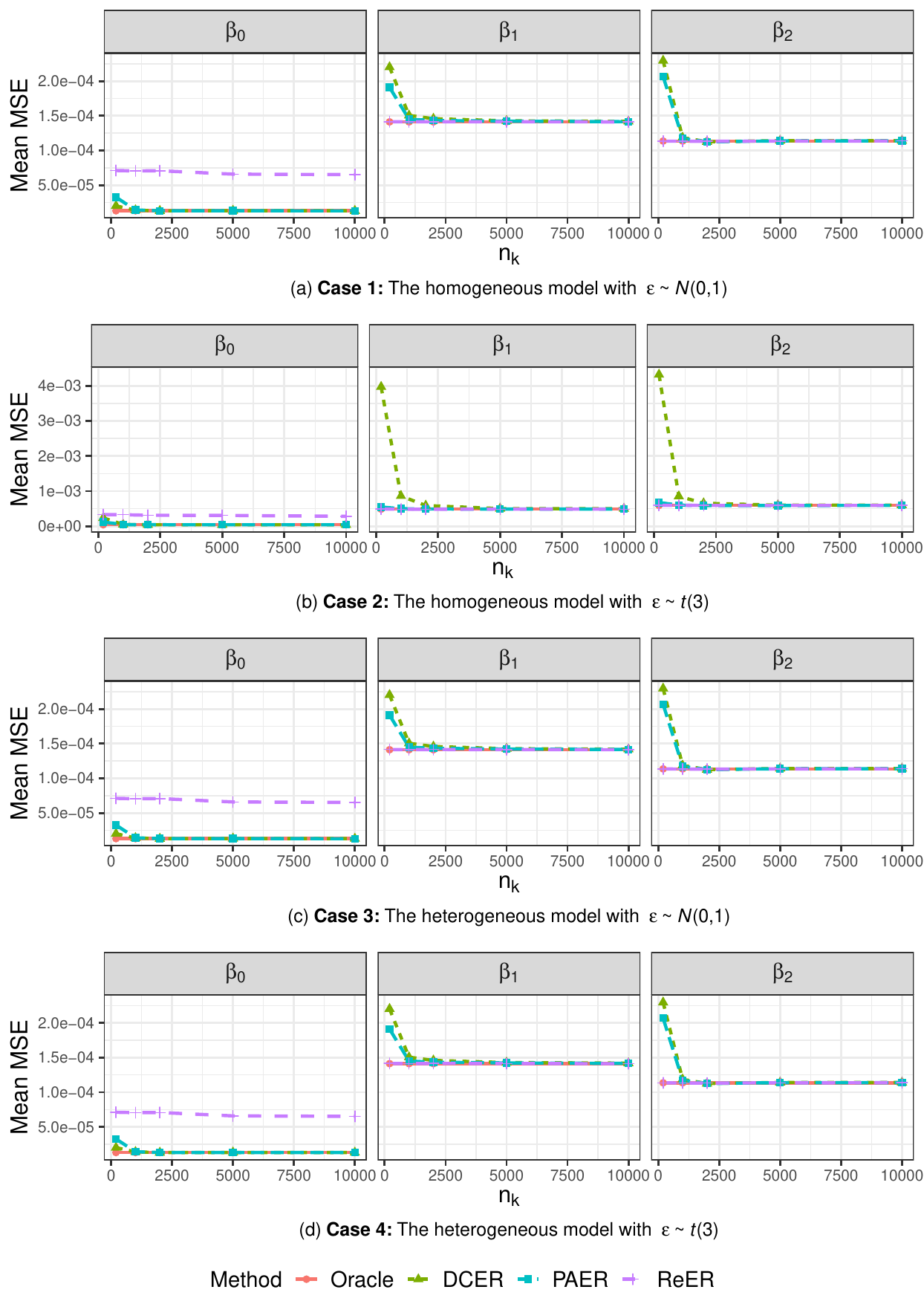}
		\caption{MSE values for fixed $N_k$ with varying $n_k$ at the $25\%$ expectile level.}
		\label{fig:S1}
	\end{figure}

	Specifically, Table~\ref{tab:fix-homo} presents the simulation results for Scenario \textbf{S1} under the homogeneous model setting. We observe the following:
	\begin{itemize}
		\item[1)] First, except for the intercept term $\beta_0$, the proposed \textbf{ReER} method achieves performance comparable to the \textbf{Oracle} estimator under both the standard normal and heavy-tailed distributions.
		\item[2)] Second, in the heavy-tailed setting, it is notable that \textbf{DCER} performs worse than the other methods when $n_k$ is small, while \textbf{ReER} and \textbf{PAER} remain comparable to the \textbf{Oracle}.
		\item[3)] Third, as $n_k$ decreases, the MSE for \textbf{DCER} and \textbf{PAER} tends to increase, whereas the proposed \textbf{ReER} remains stable (see also Figure~\ref{fig:S1} for a detailed illustration). This may be because \textbf{DCER} and \textbf{PAER} rely primarily on the data within each individual batch. When the batch size $n_k$ is small, estimation accuracy deteriorates, reducing performance. In contrast, \textbf{ReER} estimates parameters by incorporating information across batches, enabling the online estimator to effectively aggregate data over multiple batches and thus mitigate the uncertainty associated with any single batch.
	\end{itemize}
	
	\begin{table}[htbp]
		\small
		\renewcommand{\arraystretch}{0.7}
		\centering
		\caption{Simulation results for $\tau=0.25$ under the homogeneous model setting, with fixed $N_k=100,000$ and varying $n_k=200,1000,2000,5000$ and $10000$, respectively. Left panel: $\epsilon \sim N(0,1)$; Right panel: $\epsilon \sim t(3)$. Rows shown in bold indicate the best results among the three renewable methods.}
		\begin{tabular}{cc|cccccccc}
			\toprule
			\multicolumn{2}{c|}{\textbf{Error term}} & \multicolumn{4}{c}{$\epsilon \sim N(0,1)$} & \multicolumn{4}{c}{$\epsilon \sim t(3)$}\\
			\midrule    
			\multicolumn{2}{c|}{\textbf{Method}} & \textbf{Oracle} & \textbf{DCER} & \textbf{PAER} & \textbf{ReER} & \textbf{Oracle} & \textbf{DCER} & \textbf{PAER} & \textbf{ReER} \\
			\midrule
			&  & \multicolumn{8}{c}{$n_k = 200$} \\
			& Time  & 0.131  & 0.043  & 0.065  & 0.020  & 0.126  & 0.043  & 0.067  & 0.021  \\
			\multirow{2}[0]{*}{$\beta_0$} & Bias  & 0.007  & -1.670  & -4.092  & \textbf{-0.301}& -0.012  & 13.327  & -7.876  & \textbf{0.545}\\
			& MSE   & 0.013  & \textbf{0.020}& 0.033  & 0.071  & 0.045  & 0.240  & \textbf{0.123}& 0.334  \\
			\multirow{2}[0]{*}{$\beta_1$} & Bias  & 1.295  & 8.188  & 6.287  & \textbf{1.346}& 0.759  & 58.994  & 5.427  & \textbf{0.934}\\
			& MSE   & 0.141  & 0.220  & 0.191& \textbf{0.141}& 0.488  & 3.970  & 0.549  & \textbf{0.490}\\
			\multirow{2}[1]{*}{$\beta_2$} & Bias  & -0.709  & 10.679  & 9.306  & \textbf{-0.635}& -1.956  & 61.565  & 8.108  & \textbf{-1.756}\\
			& MSE   & 0.114  & 0.230  & 0.207  & \textbf{0.113}& 0.594  & 4.320  & 0.677  & \textbf{0.597}\\
			\midrule
			& & \multicolumn{8}{c}{$n_k = 1000$ } \\
			& Time  & 0.137  & 0.027  & 0.028  & 0.014& 0.132  & 0.025  & 0.030  & 0.014  \\
			\multirow{2}[0]{*}{$\beta_0$} & Bias  & 0.007  & -0.406  & -0.814  & \textbf{-0.380}& -0.012  & 3.083  & -1.552  & \textbf{0.478}\\
			& MSE   & 0.013  & \textbf{0.014}& \textbf{0.014}& 0.070  & 0.045  & 0.052& \textbf{0.048}& 0.330  \\
			\multirow{2}[0]{*}{$\beta_1$} & Bias  & 1.295  & 2.689  & 2.219  & \textbf{1.358}& 0.759  & 18.351  & 1.699  & \textbf{0.978}\\
			& MSE   & 0.141  & 0.151  & 0.145  & \textbf{0.141}& 0.488  & 0.865  & 0.501  & \textbf{0.487}\\
			\multirow{2}[1]{*}{$\beta_2$} & Bias  & -0.709  & 1.556  & 1.327  & \textbf{-0.622}& -1.956  & 17.008  & \textbf{0.056}& -1.778  \\
			& MSE   & 0.114  & 0.119  & 0.117  & \textbf{0.113}& 0.594  & 0.855  & 0.602  & \textbf{0.595}\\
			\midrule
			& & \multicolumn{8}{c}{$n_k = 2000$} \\
			& Time  & 0.132  & 0.025  & 0.025  & 0.015& 0.127  & 0.025  & 0.027  & 0.015  \\
			\multirow{2}[0]{*}{$\beta_0$} & Bias  & 0.007  & \textbf{-0.182}& -0.386  & -0.489  & -0.012  & 1.814  & -0.773  & \textbf{0.392}\\
			& MSE   & 0.013  & \textbf{0.013}& \textbf{0.013}& 0.071  & 0.045  & \textbf{0.044}& 0.045  & 0.314  \\
			\multirow{2}[0]{*}{$\beta_1$} & Bias  & 1.295  & 2.075  & 1.735  & \textbf{1.374}& 0.759  & 10.063  & 1.211  & \textbf{0.967}\\
			& MSE   & 0.141  & 0.146  & 0.143  & \textbf{0.141}& 0.488  & 0.591  & 0.494  & \textbf{0.487}\\
			\multirow{2}[1]{*}{$\beta_2$} & Bias  & -0.709  & 0.334  & \textbf{0.286}& -0.622  & -1.956  & 8.088  & \textbf{-0.920}& -1.743  \\
			& MSE   & 0.114  & \textbf{0.113}& \textbf{0.113}& \textbf{0.113}& 0.594  & 0.656  & \textbf{0.592}& 0.594  \\
			\midrule
			& & \multicolumn{8}{c}{$n_k = 5000$} \\
			& Time  & 0.130  & 0.030  & 0.030  & 0.021  & 0.132  & 0.031  & 0.029  & 0.021  \\
			\multirow{2}[0]{*}{$\beta_0$} & Bias  & 0.007  & \textbf{-0.045}& -0.164  & -0.368  & -0.012  & 0.958  & \textbf{-0.263}& 0.579  \\
			& MSE   & 0.013  & \textbf{0.013}& \textbf{0.013}& 0.066  & 0.045  & \textbf{0.044}& \textbf{0.044}& 0.311  \\
			\multirow{2}[0]{*}{$\beta_1$} & Bias  & 1.295  & 1.589  & 1.478  & \textbf{1.362}& 0.759  & 4.174  & \textbf{0.865}& 0.962  \\
			& MSE   & 0.141  & 0.143  & 0.142  & \textbf{0.141}& 0.488  & 0.504  & 0.490  & \textbf{0.488}\\
			\multirow{2}[1]{*}{$\beta_2$} & Bias  & -0.709  & -0.314  & \textbf{-0.277}& -0.657  & -1.956  & 2.178  & \textbf{-1.601}& -1.764  \\
			& MSE   & 0.114  & \textbf{0.114}& \textbf{0.114}& \textbf{0.114}& 0.594  & 0.602  & \textbf{0.592}& 0.594  \\
			\midrule
			& & \multicolumn{8}{c}{$n_k = 10000$} \\
			& Time  & 0.146  & 0.038  & 0.038  & 0.029  & 0.144  & 0.041  & 0.040  & 0.033  \\
			\multirow{2}[0]{*}{$\beta_0$} & Bias  & 0.007  & \textbf{-0.022}& -0.074  & -0.439  & -0.012  & 0.462  & \textbf{-0.152}& 0.566  \\
			& MSE   & 0.013  & \textbf{0.013}& \textbf{0.013}& 0.065  & 0.045  & \textbf{0.044}& \textbf{0.044}& 0.284  \\
			\multirow{2}[0]{*}{$\beta_1$} & Bias  & 1.295  & 1.427  & 1.359  & \textbf{1.353}& 0.759  & 2.540  & \textbf{0.871}& 0.947  \\
			& MSE   & 0.141  & 0.142  & \textbf{0.141}& \textbf{0.141}& 0.488  & 0.502  & 0.491  & \textbf{0.488}\\
			\multirow{2}[1]{*}{$\beta_2$} & Bias  & -0.709  & -0.491  & \textbf{-0.483}& -0.656  & -1.956  & \textbf{-0.020}& -1.802  & -1.772  \\
			& MSE   & 0.114 & \textbf{0.114}  & \textbf{0.114}  & \textbf{0.114}  & 0.594  & 0.604  & 0.595  & \textbf{0.594}\\
			\bottomrule
		\end{tabular}
		
		\label{tab:fix-homo}
		\captionsetup{justification=raggedright,singlelinecheck=false}
		\caption*{\footnotesize Bias and MSE are reported in units of $10^{-3}$.}
	\end{table}
	
	\begin{table}[htbp]
		\small
		\renewcommand{\arraystretch}{0.7}
		\centering
		\caption{Simulation results for $\tau=0.25$ under the heterogeneous model setting, with fixed $N_k=100,000$ and varying $n_k=200,1000,2000,5000$ and $10000$, respectively. Left panel: $\epsilon \sim N(0,1)$; Right panel: $\epsilon \sim t(3)$. Rows shown in bold indicate the best results among the three renewable methods.}
		\begin{tabular}{cc|cccccccc}
			\toprule
			\multicolumn{2}{c|}{\textbf{Error term}} & \multicolumn{4}{c}{$\epsilon \sim N(0,1)$} & \multicolumn{4}{c}{$\epsilon \sim t(3)$}\\
			\midrule    
			\multicolumn{2}{c|}{\textbf{Method}} & \textbf{Oracle} & \textbf{DCER} & \textbf{PAER} & \textbf{ReER} & \textbf{Oracle} & \textbf{DCER} & \textbf{PAER} & \textbf{ReER} \\
			\midrule
			&  & \multicolumn{8}{c}{$n_k = 200$} \\
			& Time  & 0.121 & 0.043 & 0.066 & 0.020 & 0.128 & 0.043 & 0.068 & 0.020 
			\\
			\multirow{2}[0]{*}{$\beta_0$} & Bias  & -0.104 & 4.383 & -3.816 & \textbf{0.985} & 0.240 & 26.271 & -7.396 & \textbf{-3.424}
			\\
			& MSE   & 0.012 & 0.037 & \textbf{0.031} & 0.084 & 0.048 & 0.752 & \textbf{0.121} & 0.401 
			\\
			\multirow{2}[0]{*}{$\beta_1$} & Bias  & -1.043 & -3.145 & 5.604 & \textbf{-1.126} & 3.594 & 53.098 & 11.355 & \textbf{4.220}
			\\
			& MSE   & 0.119 & \textbf{0.128} & 0.152 & 0.150 & 0.555 & 3.294 & \textbf{0.703} & 0.705 
			\\
			\multirow{2}[1]{*}{$\beta_2$} & Bias  & -0.792 & 8.119 & 9.039 & \textbf{-0.829} & 3.124 & 57.484 & 12.765 & \textbf{3.684} 
			\\
			& MSE   & 0.116 & 0.185 & 0.200 & \textbf{0.147} & 0.526 & 3.838 & 0.727 & \textbf{0.670} \\
			\midrule
			& & \multicolumn{8}{c}{$n_k = 1000$ } \\
			& Time  & 0.120 & 0.025 & 0.029 & 0.015 & 0.122 & 0.026 & 0.028 & 0.014 
			\\
			\multirow{2}[0]{*}{$\beta_0$} & Bias  & -0.104 & 0.845 & \textbf{-0.797} & 0.926 & 0.240 & 7.248 & \textbf{-1.391} & -3.502 
			\\
			& MSE   & 0.012 & 0.014 & \textbf{0.013} & 0.084 & 0.048 & 0.100 & \textbf{0.052} & 0.403 
			\\
			\multirow{2}[0]{*}{$\beta_1$} & Bias  & -1.043 & -1.586 & 0.236 & \textbf{-1.141} & 3.594 & 18.899 & 5.315 & \textbf{4.236} 
			\\
			& MSE   & 0.119 & 0.121 & \textbf{0.118} & 0.149 & 0.555 & 0.850 & \textbf{0.589} & 0.707 
			\\
			\multirow{2}[1]{*}{$\beta_2$} & Bias  & -0.792 & 0.854 & 1.149 & \textbf{-0.817} & 3.124 & 17.404 & 5.097 & \textbf{3.630} 
			\\
			& MSE   & 0.116 & \textbf{0.116} & 0.117 &  0.147 & 0.526 & 0.831 & \textbf{0.560} & 0.667 \\
			\midrule
			& & \multicolumn{8}{c}{$n_k = 2000$} \\
			& Time  & 0.125 & 0.023 & 0.025 & 0.014 & 0.126 & 0.024 & 0.024 & 0.015 
			\\
			\multirow{2}[0]{*}{$\beta_0$} & Bias  & -0.104 & \textbf{0.391} & -0.418 & 0.881 & 0.240 & 4.355 & \textbf{-0.578} & -3.659 
			\\
			& MSE   & 0.012 & \textbf{0.013} & \textbf{0.013} & 0.084 & 0.048 & 0.065 & \textbf{0.050} & 0.401 
			\\
			\multirow{2}[0]{*}{$\beta_1$} & Bias  & -1.043 & -1.285 & \textbf{-0.455} & -1.146 & 3.594 & 12.040 & 4.512 & \textbf{4.228} 
			\\
			& MSE   & 0.119 & 0.122 & \textbf{0.119} & 0.149 & 0.555 & 0.644 & \textbf{0.570 }& 0.702 
			\\
			\multirow{2}[1]{*}{$\beta_2$} & Bias  & -0.792 & \textbf{0.053} & 0.151 & -0.820 & 3.124 & 10.214 & 4.084 & \textbf{3.621} 
			\\
			& MSE   & 0.116 & \textbf{0.116} & \textbf{0.116} & 0.147 & 0.526 & 0.626 & \textbf{0.543} & 0.664 \\
			\midrule
			& & \multicolumn{8}{c}{$n_k = 5000$} \\
			& Time  & 0.126 & 0.029 & 0.029 & 0.020 & 0.127 & 0.028 & 0.029 & 0.020 
			\\
			\multirow{2}[0]{*}{$\beta_0$} & Bias  & -0.104 & 0.077 & -0.233 & 0.875 & 0.240 & 2.289 & -0.088 & -3.904 
			\\
			& MSE   & 0.012 & \textbf{0.012} & \textbf{0.012} & 0.083 & 0.048 & 0.052 & \textbf{0.050} & 0.409 
			\\
			\multirow{2}[0]{*}{$\beta_1$} & Bias  & -1.043 & -1.138 & \textbf{-0.832} & -1.138 & 3.594 & 7.169 & \textbf{3.917} & 4.229 
			\\
			& MSE   & 0.119 & 0.120 & \textbf{0.119} & 0.148 & 0.555 & 0.564 & \textbf{0.560} & 0.701 
			\\
			\multirow{2}[1]{*}{$\beta_2$} & Bias  & -0.792 & -0.466 & \textbf{-0.438} & -0.810 & 3.124 & 5.653 & 3.574 & \textbf{3.664} 
			\\
			& MSE   & 0.116 & \textbf{0.117} & \textbf{0.117} & 0.145 & 0.526 & 0.547 & \textbf{0.532} & 0.663 \\
			\midrule
			& & \multicolumn{8}{c}{$n_k = 10000$} \\
			
			& Time  & 0.134 & 0.041 & 0.040 & 0.032 & 0.137 & 0.041 & 0.044 & 0.034 
			\\
			\multirow{2}[0]{*}{$\beta_0$} & Bias  & -0.104 & \textbf{-0.015} & -0.153 & 0.890 & 0.240 & 1.299 & \textbf{0.091} & -3.722 
			\\
			& MSE   & 0.012 & \textbf{0.012} & \textbf{0.012} & 0.082 & 0.048 & \textbf{0.048} & \textbf{0.048} & 0.387 
			\\
			\multirow{2}[0]{*}{$\beta_1$} & Bias  & -1.043 & -1.075 & \textbf{-0.962} & -1.149 & 3.594 & 5.517 & \textbf{3.829} & 4.235 
			\\
			& MSE   & 0.119 & \textbf{0.119} & \textbf{0.119} & 0.147 & 0.555 & 0.562 & \textbf{0.557} & 0.690 
			\\
			\multirow{2}[1]{*}{$\beta_2$} & Bias  & -0.792 & -0.660 & \textbf{-0.642} & -0.823 & 3.124 & 4.442 & \textbf{3.316}& 3.618 
			\\
			& MSE   & 0.116 & 0.118 & \textbf{0.117} & 0.144 & 0.526 & \textbf{0.528} & 0.530 & 0.658 \\
			\bottomrule
		\end{tabular}
		
		\label{tab:fix-hetero}
		\captionsetup{justification=raggedright,singlelinecheck=false}
		\caption*{\footnotesize Bias and MSE are reported in units of $10^{-3}$.}
	\end{table}
	Tables~\ref{tab:fix-hetero} presents the simulation results under heteroscedastic settings for normal and heavy-tailed error distributions in Scenario \textbf{S1}. Correspondingly, Figure~\ref{fig:S1} displays the MSE values. The conclusions are consistent with those drawn from Tables~\ref{tab:fix-homo}. Notably, even under heteroscedasticity, the performance of \textbf{ReER} remains stable, while the estimation accuracy of competing methods decreases as the batch size $n_k$ decreases. This underscores the advantage of \textbf{ReER} in effectively leveraging information across sequential data batches.
	
	\subsubsection{Evaluation of the the number of batches on performance}\label{S2}
	
	To assess the effect of $K$ on the statistical performance of the proposed renewable estimators, we fix the batch size at $n_k = 300$ for all $k = 1, 2, 3, \ldots, K$, and vary the total number of batches $K$ within ${100, 200, 500, 1000, 2000}$. In Scenario \textbf{S2}, the data arrive in small batches at high frequency, which better reflects real-world streaming data settings.
	
	The simulation results for the homoscedastic and heteroscedastic settings under Scenario \textbf{S2} are reported in Tables~\ref{tab:stream-homo} and Tables~\ref{tab:stream-hetero}, respectively. Specifically, as the number of batches $K$ increases, the total sample size grows accordingly, and the MSE values for our proposed method as well as for the competing estimators consistently converge toward zero under the standard normal distribution. This further demonstrates the statistical consistency of the proposed method and other renewable expectile regression approaches. Moreover, in terms of both \textbf{BIAS} and \textbf{MSE}, the performance of \textbf{ReER} is generally comparable to that of the \textbf{Oracle} estimator, except for the intercept term. Although the estimation accuracy of $\boldsymbol{\beta}_0$ for \textbf{ReER} is initially slightly lower than that of the benchmark methods, its MSE steadily decreases with increasing $K$, eventually matching the performance of the \textbf{Oracle}. Finally, from Tables~\ref{tab:stream-hetero}, it can be observed that the proposed method performs well in handling heteroscedastic scenarios. The results are consistent with those for the homoscedastic setting, so redundant discussions are omitted.
	
\begin{figure}[htbp]
		\centering
		\includegraphics[width=\linewidth,]{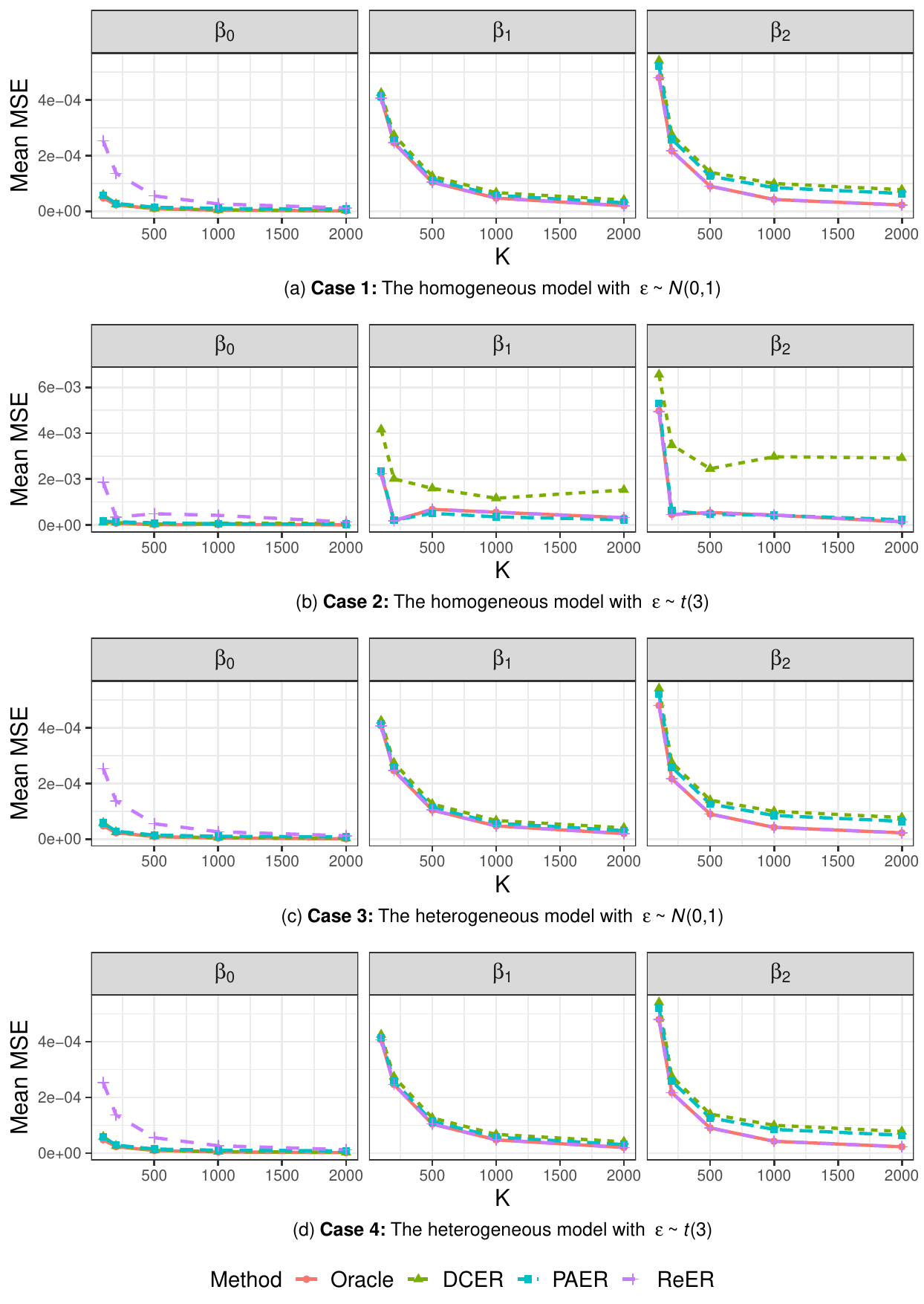}
		\caption{MSE values for fixed $n_k$ with varying $K$ at the $25\%$ expectile level.}
		\label{fig:S2}
	\end{figure}

	To visualize more intuitively, Figure~\ref{fig:S2} shows how the MSE values evolve as the batch number $K$ increases. It can be seen that, in most settings, both \textbf{ReER} and \textbf{PAER} achieve estimation accuracy comparable to that of the \textbf{Oracle}, whereas the performance of \textbf{DCER} deteriorates markedly, especially under heavy-tailed error distributions. These findings further underscore the potential robustness of the proposed \textbf{ReER} method against non-Gaussian noise. 
	
	\begin{table}[htbp]
		\small
		\renewcommand{\arraystretch}{0.7}
		\centering
		\caption{Simulation results for $\tau=0.25$ under the homogeneous model setting, with fixed $n_k$ and varying $K = 100,200,500,1000,2000$ , respectively. Left panel: $\epsilon \sim N(0,1)$; Right panel: $\epsilon \sim t(3)$. Rows shown in bold indicate the best results among the three renewable methods.}
		\begin{tabular}{cc|cccccccc}
			\toprule
			\multicolumn{2}{c|}{\textbf{Error term}} & \multicolumn{4}{c}{$\epsilon \sim N(0,1)$} & \multicolumn{4}{c}{$\epsilon \sim t(3)$}\\
			\midrule    
			\multicolumn{2}{c|}{\textbf{Method}} & \textbf{Oracle} & \textbf{DCER} & \textbf{PAER} & \textbf{ReER} & \textbf{Oracle} & \textbf{DCER} & \textbf{PAER} & \textbf{ReER} \\
			\midrule
			&  & \multicolumn{8}{c}{$K = 100$ } \\
			& Time  & 0.047 & 0.004 & 0.004 & 0.004 & 0.087 & 0.004 & 0.004 & 0.004 
			\\
			\multirow{2}[0]{*}{$\beta_0$} & Bias  & 0.879 & \textbf{-0.294} & -1.830 & 2.679 & -1.126 & 4.964 & \textbf{-3.858} & -25.786 
			\\
			& MSE   & 0.048 & 0.059 & \textbf{0.058} & 0.253 & 0.138 & \textbf{0.114} & 0.173 & 1.857 
			\\
			\multirow{2}[0]{*}{$\beta_1$} & Bias  & -2.114 & 2.721 & \textbf{1.051} & -1.930 & 18.114 & 42.486 & \textbf{16.319} & 18.389 
			\\
			& MSE   & 0.407 & 0.425 & 0.414 & \textbf{0.407} & 2.244 & 4.158 & 2.358 & \textbf{2.233}
			\\
			\multirow{2}[1]{*}{$\beta_2$} & Bias  & -1.159 & 6.910 & 5.874 & \textbf{-0.923} & 30.180 & 63.989 & \textbf{30.222} & 30.493 
			\\
			& MSE   & 0.480 & 0.541 & 0.522 & \textbf{0.480} & 4.972 & 6.552 & 5.284 & \textbf{4.943}\\
			\midrule
			& & \multicolumn{8}{c}{$K = 200$} \\
			& Time  & 0.079 & 0.004 & 0.004 & 0.004 & 0.113 & 0.004 & 0.004 & 0.004 \\
			\multirow{2}[0]{*}{$\beta_0$} & Bias  & 0.351 & \textbf{-0.682} & -2.196 & 1.201 & -0.794 & 5.790 & -5.666 & \textbf{-2.892 }
			\\
			& MSE   & 0.023 & \textbf{0.026} & 0.029 & 0.137 & 0.078 & \textbf{0.109} & 0.142 & 0.342 
			\\
			\multirow{2}[0]{*}{$\beta_1$} & Bias  & -0.778 & 4.001 & 2.491 & \textbf{-0.686} & -6.967 & 34.895 & \textbf{-4.004} & -6.773 
			\\
			& MSE   & 0.247 & 0.275 & 0.257 & \textbf{0.247} & 0.191 & 2.013 & 0.210 & \textbf{0.189} 
			\\
			\multirow{2}[1]{*}{$\beta_2$} & Bias  & -0.834 & 6.819 & 5.919 & \textbf{-0.712} & 11.583 & 57.015 & 16.510 & \textbf{11.603} 
			\\
			& MSE   & 0.217 & 0.276 & 0.259 & \textbf{0.217} & 0.458 & 3.489 & 0.618 & \textbf{0.454} \\
			\midrule
			& & \multicolumn{8}{c}{$K = 500$} \\
			& Time  & 0.197 & 0.004 & 0.004 & 0.004 & 0.149 & 0.004 & 0.004 & 0.004 
			\\
			\multirow{2}[0]{*}{$\beta_0$} & Bias  & 0.289 & \textbf{-0.746}& -2.349 & 1.024 & -1.496 & \textbf{3.908}& -8.337 & 8.753 
			\\
			& MSE   & 0.009 & \textbf{0.010} & 0.015 & 0.055 & 0.017 & \textbf{0.038} & 0.097 & 0.494 
			\\
			\multirow{2}[0]{*}{$\beta_1$} & Bias  & -0.655 & 3.980 & 2.620 & \textbf{-0.617} & -14.858 & 33.388 & \textbf{-10.672} & -14.737 
			\\
			& MSE   & 0.105 & 0.127 & 0.114 & \textbf{0.105} & 0.683 & 1.596 & \textbf{0.512} & 0.686 
			\\
			\multirow{2}[1]{*}{$\beta_2$} & Bias  & -0.750 & 6.738 & 5.934 & \textbf{-0.702} & -7.108 & 43.835 & -1.803 & \textbf{-7.130} 
			\\
			& MSE   & 0.091 & 0.140 & 0.127 & \textbf{0.090} & 0.545 & 2.452 & 0.471 & \textbf{0.546} \\
			\midrule
			& & \multicolumn{8}{c}{$K = 1000$} \\
			& Time  & 0.363 & 0.004 & 0.004 & 0.004 & 0.346 & 0.004 & 0.004 & 0.004 
			\\
			\multirow{2}[0]{*}{$\beta_0$} & Bias  & 0.149 & -0.934 & -2.454 & \textbf{0.358} & 0.115 & 6.589 & \textbf{-6.397} & 7.667 
			\\
			& MSE   & 0.004 & 0.005 & \textbf{0.010} & 0.027 & 0.022 & 0.076 & \textbf{0.057} & 0.424 
			\\
			\multirow{2}[0]{*}{$\beta_1$} & Bias  & -0.322 & 4.364 & 2.995 & \textbf{-0.303} & -17.173 & 29.575 & \textbf{-13.229} & -17.126 
			\\
			& MSE   & 0.048 & 0.067 & 0.057 & \textbf{0.048} & 0.558 & 1.168 & \textbf{0.359} & 0.559 
			\\
			\multirow{2}[1]{*}{$\beta_2$} & Bias  & -0.121 & 7.400 & 6.432 & \textbf{-0.097} & 1.501 & 51.106 & 6.922 & \textbf{1.482} 
			\\
			& MSE   & 0.043 & 0.100 & 0.085 & \textbf{0.043} & 0.433 & 2.976 & \textbf{0.427} & 0.435 \\
			\midrule
			& & \multicolumn{8}{c}{$K = 2000$} \\
			& Time  & 0.901 & 0.004 & 0.004 & 0.004 & 0.882 & 0.004 & 0.004 & 0.004 
			\\
			\multirow{2}[0]{*}{$\beta_0$} & Bias  & 0.094 & \textbf{-1.028} & -2.536 & 0.263 & 1.225 & 8.169 & -4.957 & \textbf{2.773} 
			\\
			& MSE   & 0.002 & \textbf{0.003} & 0.008 & 0.012 & 0.010 & 0.078 & \textbf{0.035} & 0.137 
			\\
			\multirow{2}[0]{*}{$\beta_1$} & Bias  & -0.098 & 4.470 & 3.164 & \textbf{-0.088} & -9.191 & 36.686 & \textbf{-5.611} & -9.164 
			\\
			& MSE   & 0.020 & 0.041 & 0.031 & \textbf{0.020} & 0.321 & 1.529 &\textbf{0.230} & 0.321 
			\\
			\multirow{2}[1]{*}{$\beta_2$} & Bias  & -0.220 & 7.345 & 6.379 & \textbf{-0.208} & 6.162 & 53.375 & 11.804 & \textbf{6.154} 
			\\
			& MSE   & 0.023 & 0.078 & 0.064 & \textbf{0.023} & 0.141 & 2.918 & 0.232 & \textbf{0.141} \\
			\bottomrule
		\end{tabular}
		
		\label{tab:stream-homo}
		\captionsetup{justification=raggedright,singlelinecheck=false}
		\caption*{\footnotesize Bias and MSE are reported in units of $10^{-3}$.}
	\end{table}
	
	\begin{table}[htbp]
		\small
		\renewcommand{\arraystretch}{0.7}
		\centering
		\caption{Simulation results for $\tau=0.25$ under the heterogeneous model setting, with fixed $n_k$ and varying $K = 100,200,500,1000,2000$ , respectively. Left panel: $\epsilon \sim N(0,1)$; Right panel: $\epsilon \sim t(3)$. Rows shown in bold indicate the best results among the three renewable methods.}
		\begin{tabular}{cc|cccccccc}
			\toprule
			\multicolumn{2}{c|}{\textbf{Error term}} & \multicolumn{4}{c}{$\epsilon \sim N(0,1)$} & \multicolumn{4}{c}{$\epsilon \sim t(3)$}\\
			\midrule    
			\multicolumn{2}{c|}{\textbf{Method}} & \textbf{Oracle} & \textbf{DCER} & \textbf{PAER} & \textbf{ReER} & \textbf{Oracle} & \textbf{DCER} & \textbf{PAER} & \textbf{ReER} \\
			\midrule
			&  & \multicolumn{8}{c}{$K = 100$ } \\
			& Time  & 0.045 & 0.004 & 0.004 & 0.004 & 0.039 & 0.004 & 0.004 & 0.004 
			\\
			\multirow{2}[0]{*}{$\beta_0$} & Bias  & 0.420 & 3.604 & -1.829 & \textbf{-0.932} & 0.464 & 18.941 & -5.311 & \textbf{-3.666} 
			\\
			& MSE   & 0.060 & 0.082 & \textbf{0.071} & 0.373 & 0.169 & 0.571 & \textbf{0.245} & 1.192 
			\\
			\multirow{2}[0]{*}{$\beta_1$} & Bias  & 1.941 & \textbf{0.349} & 5.938 & 2.349 & 4.391 & 42.949 & 9.794 & \textbf{5.308} 
			\\
			& MSE   & 0.463 & \textbf{0.478} & 0.507 & 0.586 & 1.984 & 3.457 & \textbf{2.091} & 2.511 
			\\
			\multirow{2}[1]{*}{$\beta_2$} & Bias  & 1.248 & 6.633 & 7.567 & \textbf{1.618} & 2.570 & 43.045 & 9.631 & \textbf{3.473} 
			\\
			& MSE   & 0.488 & \textbf{0.559} & 0.576 & 0.614 & 1.691 & 3.400 & \textbf{1.870} & 2.145 \\
			\midrule
			& & \multicolumn{8}{c}{$K = 200$} \\
			& Time  & 0.084 & 0.004 & 0.004 & 0.004 & 0.080 & 0.004 & 0.004 & 0.004 
			\\
			\multirow{2}[0]{*}{$\beta_0$} & Bias  & -0.043 & 3.075 & -2.425 & \textbf{-1.444} & -0.023 & 18.514 & -5.469 & \textbf{-4.014} 
			\\
			& MSE   & 0.027 & 0.044 & \textbf{0.038} & 0.175 & 0.092 & 0.445 & \textbf{0.146} & 0.630 
			\\
			\multirow{2}[0]{*}{$\beta_1$} & Bias  & 1.950 & 0.548 & 6.384 & \textbf{2.286} & 2.392 & 42.156 & 7.341 & \textbf{2.892} 
			\\
			& MSE   & 0.267 & \textbf{0.277} & 0.313 & 0.338 & 0.972 & 2.566 & \textbf{1.036} & 1.228 
			\\
			\multirow{2}[1]{*}{$\beta_2$} & Bias  & 0.987 & 6.732 & 7.580 & \textbf{1.213} & 4.369 & 44.970 & 11.554 & \textbf{5.226} 
			\\
			& MSE   & 0.218 & \textbf{0.274} & 0.285 & 0.275 & 0.823 & 2.855 & \textbf{0.975} & 1.045 \\
			\midrule
			& & \multicolumn{8}{c}{$K = 500$} \\
			& Time  & 0.191 & 0.004 & 0.004 & 0.004 & 0.199 & 0.004 & 0.004 & 0.004 
			\\
			\multirow{2}[0]{*}{$\beta_0$} & Bias  & 0.170 & 3.319 & -2.305 & \textbf{0.029} & 0.597 & 18.765 & -4.846 & \textbf{0.793} 
			\\
			& MSE   & 0.010 & 0.023 & \textbf{0.017} & 0.058 & 0.028 & 0.389 & \textbf{0.059} & 0.255 
			\\
			\multirow{2}[0]{*}{$\beta_1$} & Bias  & 0.461 & -1.056 & 4.965 & \textbf{0.552} & -0.618 & 40.001 & 4.554 & \textbf{-0.63}4 
			\\
			& MSE   & 0.099 & \textbf{0.106} & 0.125 & 0.125 & 0.354 & 1.922 & \textbf{0.383} & 0.450 
			\\
			\multirow{2}[1]{*}{$\beta_2$} & Bias  & -0.099 & 5.570 & 6.528 & \textbf{-0.059} & 0.174 & 42.753 & 6.914 & \textbf{0.320} 
			\\
			& MSE   & 0.071 & 0.105 & 0.117 & \textbf{0.090} & 0.365 & 2.175 & \textbf{0.433} & 0.462 \\
			\midrule
			& & \multicolumn{8}{c}{$K = 1000$} \\
			& Time  & 0.354 & 0.004 & 0.004 & 0.004 & 0.363 & 0.004 & 0.004 & 0.004 
			\\
			\multirow{2}[0]{*}{$\beta_0$} & Bias  & 0.139 & 3.303 & -2.340 & \textbf{0.039} & 0.163 & 18.400 & -5.217 & \textbf{0.400} 
			\\
			& MSE   & 0.004 & 0.016 & \textbf{0.010} & 0.038 & 0.014 & 0.355 & \textbf{0.045} & 0.124 
			\\
			\multirow{2}[0]{*}{$\beta_1$} & Bias  & 0.279 & -1.414 & 4.664 & \textbf{0.329} & -0.266 & 39.403 & 4.881 & \textbf{-0.275} 
			\\
			& MSE   & 0.056 & 0.058 & 0.078 & \textbf{0.071} & 0.165 & 1.704 & \textbf{0.198} & 0.210 
			\\
			\multirow{2}[1]{*}{$\beta_2$} & Bias  & -0.006 & 5.739 & 6.713 & \textbf{0.020} & -0.194 & 42.297 & 6.593 & \textbf{-0.166} 
			\\
			& MSE   & 0.044 & 0.080 & 0.091 & \textbf{0.056} & 0.179 & 1.960 & 0.230 & \textbf{0.226} \\
			\midrule
			& & \multicolumn{8}{c}{$K = 2000$} \\
			& Time  & 0.892 & 0.004 & 0.004 & 0.004 & 0.902 & 0.004 & 0.004 & 0.004 
			\\
			\multirow{2}[0]{*}{$\beta_0$} & Bias  & 0.032 & 3.132 & -2.501 & \textbf{0.005} & -0.204 & 18.065 & -5.529 & \textbf{-0.615} 
			\\
			& MSE   & 0.002 & 0.012 & \textbf{0.008} & 0.017 & 0.006 & 0.334 & \textbf{0.039} & 0.063 
			\\
			\multirow{2}[0]{*}{$\beta_1$} & Bias  & 0.132 & -1.541 & 4.551 & \textbf{0.158} & 0.765 & 40.179 & 5.825 & \textbf{0.881} 
			\\
			& MSE   & 0.023 & \textbf{0.026} & 0.044 & 0.029 & 0.081 & 1.691 & 0.120 & \textbf{0.102} 
			\\
			\multirow{2}[1]{*}{$\beta_2$} & Bias  & -0.066 & 5.652 & 6.658 & \textbf{-0.059} & -0.034 & 42.342 & 6.753 & \textbf{-0.023} 
			\\
			& MSE   & 0.021 & 0.053 & 0.065 & \textbf{0.027} & 0.078 & 1.868 & 0.131 & \textbf{0.098} \\
			\bottomrule
		\end{tabular}
		\label{tab:stream-hetero}
		\captionsetup{justification=raggedright,singlelinecheck=false}
		\caption*{\footnotesize Bias and MSE are reported in units of $10^{-3}$.}
	\end{table}
	
	\subsection{Computation time Evaluation}\label{sub:3-3}
	
\begin{figure}[ht]
		\centering
		\includegraphics[width = \linewidth]{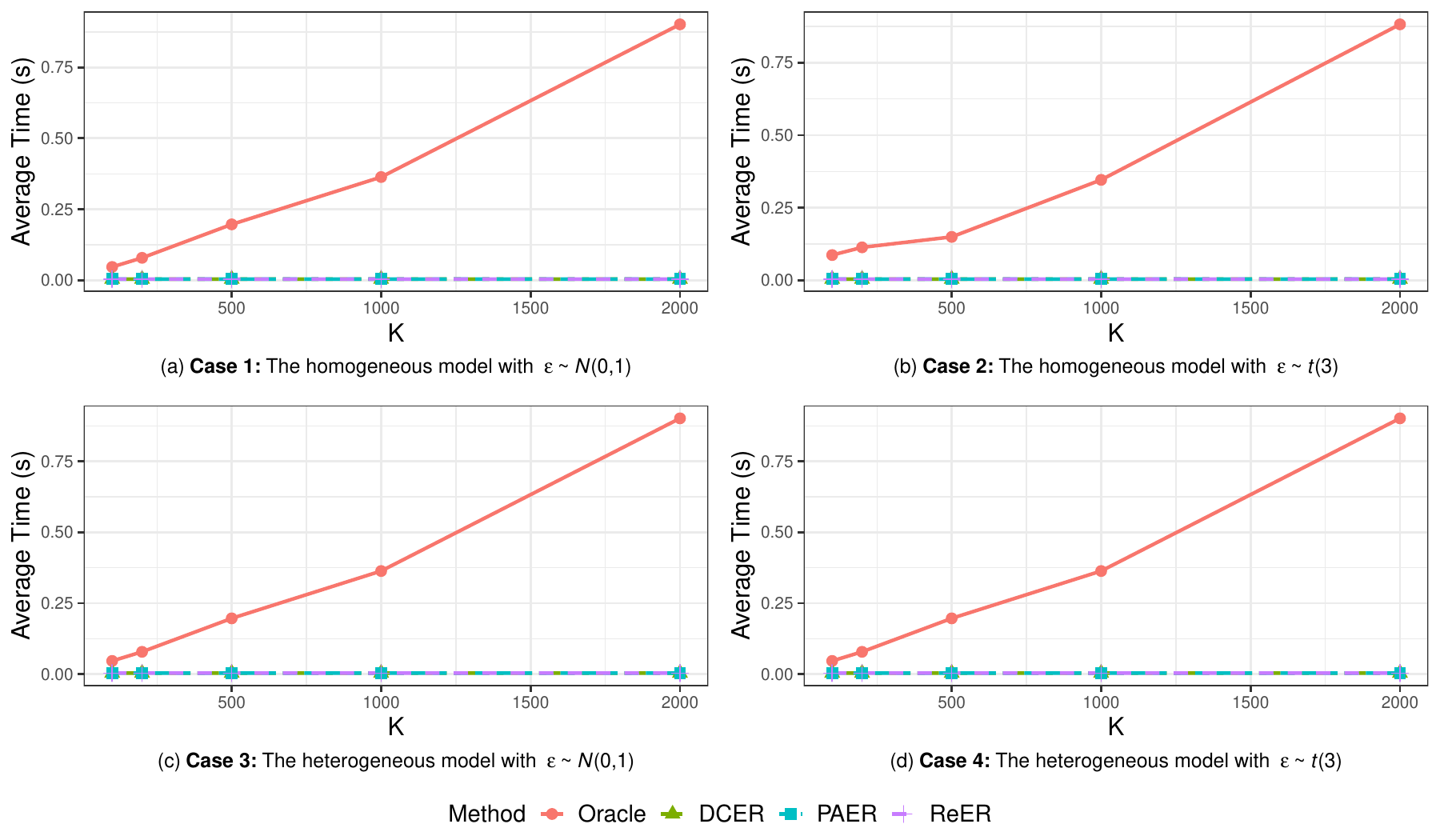}
		\caption{The computational time of renewable method and the Oracle method with fixed $n_k$ and varying $K$ in 25\% expectile level.}
		\label{fig:time}
	\end{figure}

	To evaluate the computational efficiency of the proposed renewable method, we report the average computation time over 200 replications under Scenario \textbf{S2}, as shown in Figure~\ref{fig:time}. As expected, all three online learning methods consistently demonstrate lower computation time than the full-data-based \textbf{Oracle} estimator, highlighting the computational advantage of online learning approaches for large-scale data analysis.
	
	To further compare computational efficiency among the renewable estimators, Table~\ref{tab:time-improve} reports the average time for each of the three renewable methods, along with the average percentage difference in computation time between \textbf{ReER} and the two benchmark methods (\textbf{DCER} and \textbf{PAER}) (shown in parentheses). It can be observed that \textbf{ReER} achieves the lowest computation time, with an approximate 2\% reduction compared to both \textbf{DCER} and \textbf{PAER}. This efficiency gain is likely due to the design of \textbf{ReER}, which updates parameter estimates without performing a full expectile regression for each new data batch, thereby avoiding repeated batch-level optimization and reducing overall computation time.
	
	\begin{table}[t]
		\renewcommand{\arraystretch}{0.7}
		\centering
		\caption{Average computational time (in units of $10^{-3}$) of the renewable methods, with percentage differences relative to \textbf{ReER} presented in parentheses.}
		\begin{tabular}{ccccc}
			\toprule
			\textbf{Method} & \textbf{Case 1} & \textbf{Case 2} & \textbf{Case 3} & \textbf{Case 4} \\
			\midrule
			& \multicolumn{4}{c}{$K = 100$} \\
			\cmidrule{2-5}    \textbf{DCER} & 4.00(-1.1\%)& 3.82(-1.2\%)& 4.02(-1.3\%)& 4.02(-1.7\%)\\
			\textbf{PAER} & 4.08(-3.0\%)& 3.86 (-2.2\%)& 4.06 (-2.2\%)& 4.08 (-3.1\%)\\
			\textbf{ReER}& 3.96& 3.78& 3.97&3.95\\
			\midrule
			& \multicolumn{4}{c}{$K = 200$} \\
			\cmidrule{2-5}    \textbf{DCER} & 3.96 (-1.5\%)& 3.91 (-1.2\%)& 3.90 (-1.4\%)& 3.94 (-1.3\%)\\
			\textbf{PAER} & 4.01 (-2.7\%)& 3.95 (-2.2\%)& 3.93 (-2.4\%)& 3.98 (-2.3\%)\\
			\textbf{ReER}& 3.90& 3.86& 3.84&3.89\\
			\midrule
			& \multicolumn{4}{c}{$K = 500$} \\
			\cmidrule{2-5}    \textbf{DCER} & 3.95 (-1.3\%)& 3.93 (-1.2\%)& 3.92 (-1.3\%)& 3.93 (-1.3\%)\\
			\textbf{PAER} & 3.99 (-2.4\%)& 3.98 (-2.4\%)& 3.96 (-2.4\%)& 3.97 (-2.4\%)\\
			\textbf{ReER}& 3.90& 3.89& 3.87&3.87\\
			\midrule
			& \multicolumn{4}{c}{$K = 1000$} \\
			\cmidrule{2-5}    \textbf{DCER} & 3.91 (-1.3\%)& 3.89 (-1.3\%)& 3.90 (-1.3\%)& 3.88 (-1.3\%)\\
			\textbf{PAER} & 3.95 (-2.4\%)& 3.93 (-2.3\%)& 3.94 (-2.3\%)& 3.92 (-2.3\%)\\
			\textbf{ReER}& 3.86& 3.84& 3.85&3.83\\
			\midrule
			& \multicolumn{4}{c}{$K = 2000$} \\
			\cmidrule{2-5}    \textbf{DCER} & 3.93 (-1.3\%)& 3.92 (-1.1\%)& 3.90 (-1.3\%)& 3.91 (-1.3\%)\\
			\textbf{PAER} & 3.97 (-2.3\%)& 3.97 (-2.2\%)& 3.94 (-2.4\%)& 3.95 (-2.3\%)\\
			\textbf{ReER}& 3.87& 3.88& 3.85&3.86\\
			\bottomrule
			& & & &\\
		\end{tabular}
		\label{tab:time-improve}
		\captionsetup{justification=raggedright,singlelinecheck=false}
	\end{table}
	
	\section{Empirical study}\label{sec:realdata}
	
	For illustration purpose, we apply the proposed method to analyze data sets from two practical examples in this section.

	\subsection{Beijing multisite air-quality dataset}\label{sub:4-1}
	The Beijing multi-site air-quality dataset contains 420,768 hourly measurements of air pollutants collected from 12 nationally controlled air quality monitoring stations across Beijing. This  data were provided by the Beijing Municipal Environmental Monitoring Center and cover the period from January 1, 2013, to December 31, 2014. The dataset is publicly available through the UCI Machine Learning Repository (\href{https://archive.ics.uci.edu/ml/datasets/Beijing+Multi-Site+Air-Quality+Data}{Air Quality Dataset}). The aim of this real-world case study is to explore the relationship between $\mathrm{PM}_{2.5}$ ($\mu\mathrm{g}/\mathrm{m}^3$) and other covariates, which are described in Table~\ref{tab:air-var}. Since the observations come from 12 different monitoring sites, we treat these as 12 separate data streams, setting the number of data batches to $b = 12$.
	
	\begin{table}[htbp]
		\centering
		\caption{Covariates and descriptions for air quality dataset.}
		\renewcommand{\arraystretch}{0.8}
		\begin{tabular}{cc}
			\toprule
			\textbf{Variable} & \textbf{Description} \\
			\midrule
			$\mathrm{SO}_2$   & $\mathrm{SO}_2$ concentration ($\mu\mathrm{g}/\mathrm{m}^3$) \\
			$\mathrm{NO}_2$   & $\mathrm{SO}_2$ concentration ($\mu\mathrm{g}/\mathrm{m}^3$) \\
			CO    & CO concentration ($\mu\mathrm{g}/\mathrm{m}^3$) \\
			TEMP  & temperature (degrees Celsius) \\
			PRES  & pressure (hPa) \\
			DEWP  & dew point temperature (degrees Celsius) \\
			WSPM  & wind speed (m/s) \\
			\bottomrule
		\end{tabular}
		\label{tab:air-var}
	\end{table}
	
	We exclude all observations with missing values, resulting in a final sample size of 388,817. To evaluate the predictive performance of the proposed \textbf{ReER} method, we use data from the first 11 batches as the training set and reserve the 12th batch for testing.  Model accuracy is assessed using the Mean Expectile Regression Prediction Error (\textbf{MPE}) defined as
	$$\mathrm{MPE} = \frac{1}{n} \sum_{i=1}^n \rho_{\tau} (y_i - \hat{y}_i(\tau)),$$ 
	where $\hat{y}_i (\tau)$ is the predicted value at $\tau$ expectile level of the testing sample. Table~\ref{tab:air} summarizes the prediction errors and computational costs for the different estimation methods. Due to the large magnitude of the response variable, the MPEs are presented in units of $10^3$ for readability. It can be observed that the three online estimation procedures achieve prediction performance comparable to that of the \textbf{Oracle} method, while significantly reducing computational cost.
	
	\begin{table}[htbp]
		\centering
		\renewcommand{\arraystretch}{0.8}
		\caption{The MPEs (reported in units of $10^{3}$) and Time (average computation times in seconds) with $\tau = 0.2,0.5,0.8$ for the air quality dataset.}
		\begin{tabular}{ccccc}
			\toprule
			\multirow{2}[4]{*}{\textbf{Method}} & \multirow{2}[4]{*}{Time} & \multicolumn{3}{c}{MPE} \\
			\cmidrule{3-5}          &       & 0.2   & 0.5   & 0.8 \\
			\midrule
			\textbf{Oracle} & 9.38921 & 0.700 & 1.090 & 0.974 \\
			\textbf{DCER} & 0.939 & 0.700 & 1.094 & 0.983 \\
			\textbf{PAER} & 0.892 & 0.697 & 1.090 & 0.978 \\
			\textbf{ReER} & 0.973 & 0.700 & 1.090 & 0.975 \\
			\bottomrule
		\end{tabular}
		\label{tab:air}
		\captionsetup{justification=raggedright,singlelinecheck=false}
	\end{table}
	
\begin{figure}[ht]
		\centering
		\includegraphics[width=1\linewidth]{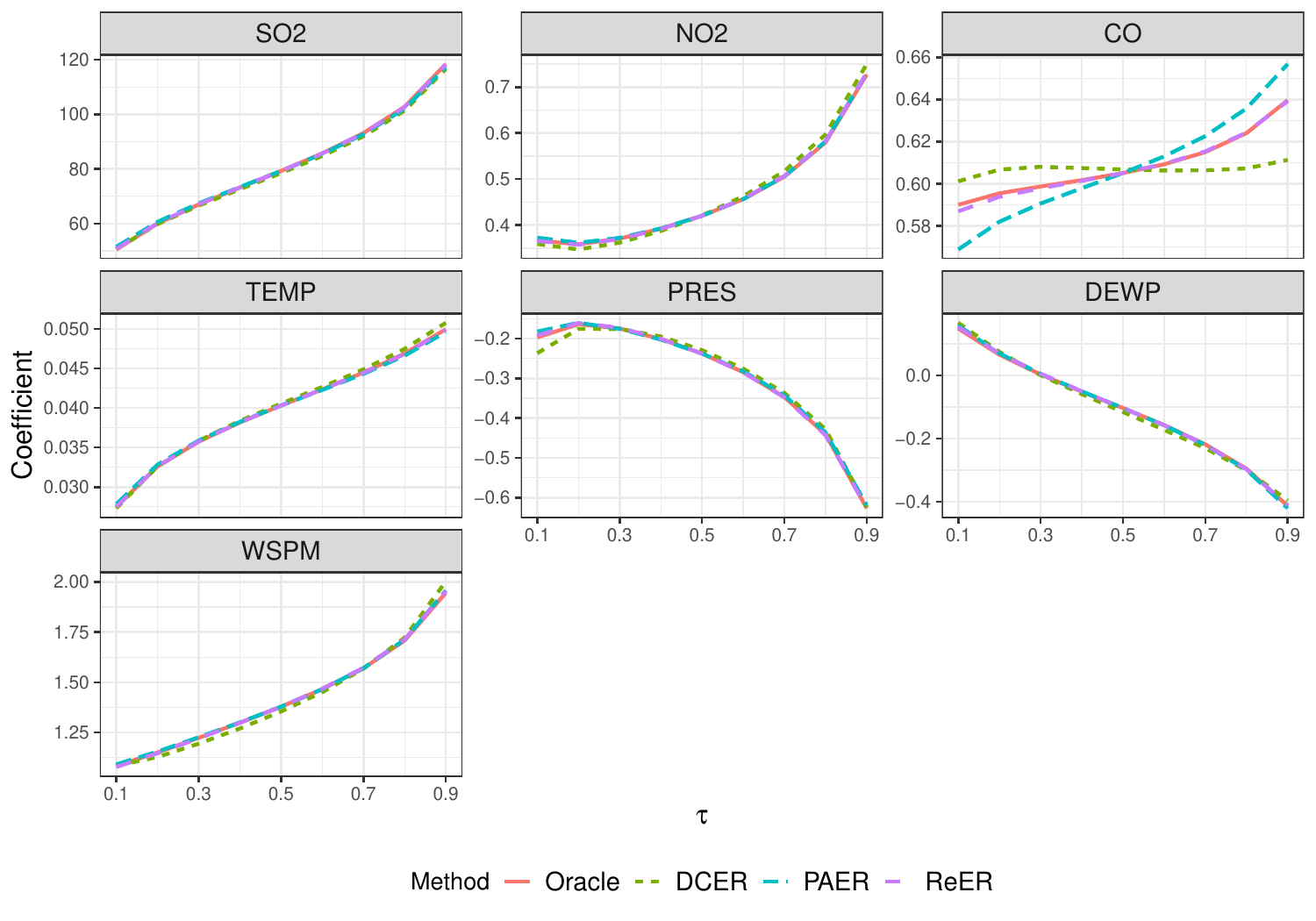}
		\caption{The estimated coefficients under different expectile level  for air quality data.}
		\label{fig:air-beta}
	\end{figure}

	To further assess the performance of the different methods, we plot the estimated coefficients across expectile levels $\tau \in \{0.1, 0.2, \ldots, 0.9\}$, as shown in Figure~\ref{fig:air-beta}. Figure~\ref{fig:air-beta} illustrates the estimated coefficients for each covariate at various expectile levels. In general, the parameter estimates obtained by the three online updating methods are largely consistent with those from the \textbf{Oracle} estimator. However, for the $\mathrm{CO}$ variable, the estimates from \textbf{DCER} and \textbf{PAER} show noticeable deviations from the \textbf{Oracle} at extreme $\tau$ values, whereas \textbf{ReER} remains closer. Since the true parameter values are unknown in real-world settings, it is not possible to directly determine which method is most accurate. Nonetheless, the fact that \textbf{ReER} produces estimates more similar to the \textbf{Oracle}, which uses the entire dataset, further demonstrates the effectiveness of the proposed approach. This suggests that \textbf{ReER} more effectively leverages information across data batches and better approximates full-sample estimation results.
	
	\subsection{Electricity consumption dataset}\label{sub:4-2}
	
	To further evaluate the performance of the proposed method we apply \textbf{ReER} to the Household Electric Power dataset (\href{https://archive.ics.uci.edu/ml/datasets/Individual+ household+electric+power+consumption}{UCI/electricity}). This dataset records electric power consumption in a single household at one-minute intervals between December 2006 and November 2010. After removing samples with missing values, the total number of observations is 2,049,280. The Household Electric Power dataset includes two key variables: \textit{Active power}, which represents the actual power consumed, and \textit{Reactive power}, which refers to the electrical power involved in energy exchange within the circuit. The aim of this application is to predict active power based on reactive power across different expectile levels. To assess predictive performance, we split the first 2,000,000 observations as the training set and use the remaining data as the testing set. Performance is also evaluated by the mean expectile regression prediction error. To further investigate the effect of the number of batches on the proposed method’s performance, we vary the number of batches  $k\in\{10,20,50,100,500\}$. Consequently, the batch size decreases from $20,000$ to $400$.
	
\begin{figure}[ht]
		\centering
		\includegraphics[width=1\linewidth]{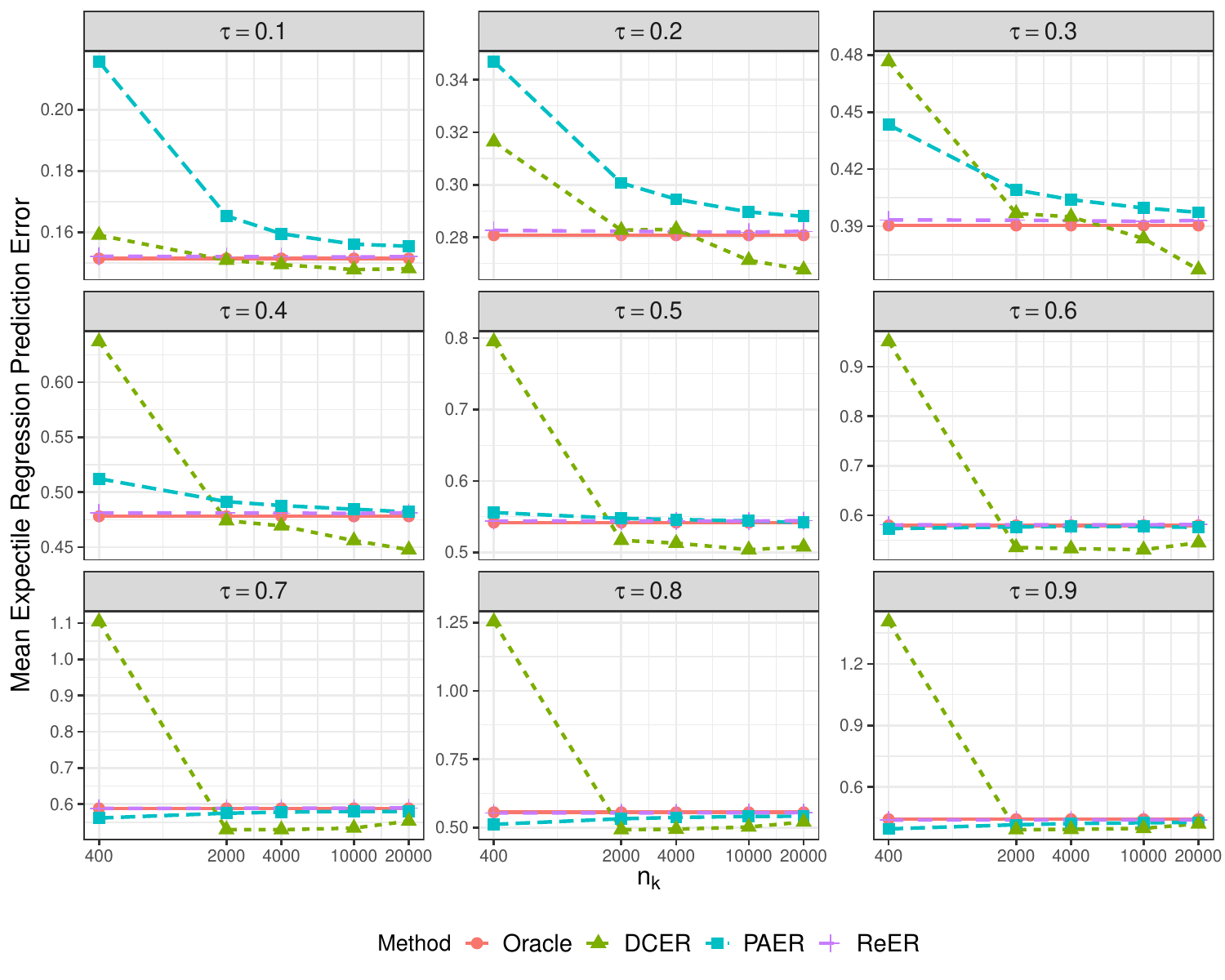}
		\caption{MPEs of the \textbf{Oracle}, \textbf{ReER}, \textbf{PAER} and \textbf{DCER} estimators for varying $n_k$ at expectile levels $\tau=0.1,0.2,\cdots,0.9$.}
		\label{fig:ele}
	\end{figure}

	Figure~\ref{fig:ele} shows the MPEs of different estimators at different expectile levels with varying $n_k$. From this figure, we observe that as the sample size $n_k$ changes, the MPEs of the \textbf{DCER} and \textbf{PAER} methods exhibits significant variation, whereas the MPEs of the \textbf{ReER} method remains nearly unchanged and is close to that of the \textbf{Oracle} method. This indicates that the proposed \textbf{ReER} method is less sensitive to the sample size of each batch. It also further demonstrates the advantage of \textbf{ReER} in effectively utilizing information across data batches for parameter estimation and prediction.
	
	\section{Conclusion and Discussion}
	\label{sec:discussion}
	
	In this paper, we propose a novel online renewable expectile regression method, \textbf{ReER}, for analyzing streaming datasets. The renewable estimator relies only on the current data and summary statistics from historical data, making it computationally efficient and free from storage constraints. We establish theoretical results under mild conditions, showing that the proposed estimator is asymptotically equivalent to the oracle expectile regression estimator. Both simulation studies and a real data application demonstrate that the proposed method achieves estimation performance comparable to the oracle estimator and superior to existing methods. In particular, unlike previous approaches, \textbf{ReER} does not update estimates simply by reweighting information from independent data streams; instead, it refines historical parameter estimates using newly arriving data, thereby more fully utilizing information across batches. As a result, even with small-sample data streams, \textbf{ReER} can better approximate oracle results and is less sensitive to batch size, effectively mitigating large estimation errors that may occur with unstable estimates from small batches. The proposed algorithm has been implemented into an R function \href{https://github.com/Weiccao/ReER}{ReER}.
	
	To conclude this paper, we discuss some limitations of this work and suggest potential directions for future research. First, the proposed \textbf{ReER} method primarily addresses online updating in low-dimensional settings. However, in the era of big data, extremely high-dimensional datasets have become increasingly common. To date, only a few studies have tackled the challenges of modeling high-dimensional streaming data. Some penalty-based or subsampling approaches have been developed to construct expectile regression models for large-scale data \citep{pan2021distributed, li2024poisson, chen2024estimation}, but they do not address online updating of high-dimensional expectile regression parameters in streaming scenarios. Therefore, developing efficient and accurate online updating methods for high-dimensional expectile regression, along with corresponding statistical inference, represents a valuable area for future research.
	
	Second, although expectile regression is an effective tool for capturing heterogeneity within data structures, challenges persist when dealing with model heterogeneity, especially in streaming data contexts. Some recent studies have explored approaches for handling heterogeneity or dynamic model shifts in streaming datasets \citep{wei2023adaptive, luo2023multivariate}, but these primarily focus on mean regression, which overlooks the heterogeneity captured by expectiles. \citet{chenRenewableQuantileRegression2024} proposed a homogeneity testing method for renewable quantile regression; however, as discussed earlier, quantile regression faces computational complexity issues. Thus, addressing model heterogeneity for expectile regression in streaming data remains an important and promising direction for future work.
	
	\section*{Acknowledgments}
	
	This research was financially supported by the State Key Program of National Natural Science Foundation of China [Nos. 72531002].
	
	\section*{Disclosure statement}
	
	No potential conflict of interest was reported by the author(s).

	\section*{Data availability}
	
	The data sets analyzed in the current study are included in the article.  Air quality dataset is publicly available through the UCI Machine Learning Repository (\href{https://archive.ics.uci.edu/ml/datasets/Beijing+Multi-Site+Air-Quality+Data}{Air Quality Dataset}), and the Household Electric Power dataset (\href{https://archive.ics.uci.edu/ml/datasets/Individual+ household+electric+power+consumption}{UCI/electricity}).
	
	\bibliography{mybib}
	\bibliographystyle{apalike}
	
	\newpage
	\appendix  
	
	\section{Technical Proofs} \label{app:proof}
	\subsection{Lemma \ref{L1}. and technical proofs}\label{proof-sub1}
	\begin{lemma}\label{L1}
		Assume that Conditions (C1)-(C5) hold, then we have:
		$$\frac{1}{\nb}\left\|W_{\nb}(\renew_b)-W_{\nb}(\truebeta)\right\|=O_p(\|\renew_b-\truebeta\|).$$
	\end{lemma}
	
	Technical proofs of Lemma~\ref{L1} are provided as follows. According to~\eqref{eq:IWLS-1} we have
	\begin{align*}
		\frac{1}{\nb}\left(W_{\nb}(\renew_b)-W_{\nb}(\truebeta)\right) 
		&= \frac{1}{\nb}\sum_{i=1}^{n_b} \left[\left| \tau - I(y_{bi} < \x_{bi}^\top \renew_b)\right| - \left|\tau - I(y_{bi} < \x_{bi}^\top \truebeta)\right| \right] \x_{bi} \x_{bi}^\top.
	\end{align*}
	Denote $\Psi_{bi}= \left[|\tau - I(y_{bi} < \x_{bi}^\top \renew_b)| - |\tau - I(y_{bi} < \x_{bi}^\top \truebeta)| \right]$. Since $I(\cdot)$ is an indicator function, which equals to $0$ or $1$. It is easy to verify that $\Psi_{bi}(\bbeta) \neq 0$ only when 
	$| \x_{bi}^\top (\renew_b - \truebeta) | \geq | y_{bi} - \x_{bi}^\top \bbeta_0 |$. Thus, it holds
	\begin{align*}
		\mathbb{E}\left[\Psi_{bi}(\bbeta)\right]&=2\cdot\left((1-2\tau)\cdot\left[F_\e\left(\x_{bi}^\top(\renew_b-\truebeta)\right)-F_\e(0)\right]\right) \\
		&=2\cdot(1-2\tau)\cdot f_\e(\xi_{bi})\cdot\left(\x_{bi}^\top(\renew_b-\truebeta)\right),
	\end{align*}
	where $\xi_{bi}$ lies between $0$ and $\x_{bi}^\top(\renew_b-\truebeta)$. 
	Under (C4) and (C5) we have
	\begin{align*}
		\left\|\mathbb{E}\left[\Psi_{bi}(\bbeta)\right]\right\|&=\|2(2\tau-1)\cdot f_\e(\xi_{bi})\cdot\x_{bi}^\top(\renew_b-\truebeta)\| \leq c_5\cdot\|\renew_b-\truebeta\|,
	\end{align*}
	where $c_5=2|2\tau-1|\cdot c_3 \cdot c_4$. Then with (C3), we have
	\begin{align*}
		\frac{1}{\nb}\left\|W_{\nb}(\renew_b)-W_{\nb}(\truebeta)\right\|&\leq c_6\cdot \|\renew_b-\truebeta\|=O_p(\|\renew_b-\truebeta\|),
	\end{align*}
	where $c_6=c_5 \cdot \left\|\frac{1}{\nb} \sum_{i=1}^{n_b}\x_{bi} \x_{bi}^\top \right\|$. As desired.
	
	\subsection{Technical proofs of Theorem~\ref{T1}.}\label{proof-sub2}
	Let $\truebeta$ be the true parameter, for the initial estimator, we have $\hatbeta_1=\renew_1$, which  denote the static estimator and renewable  estimator, receptively. Under (C1)-(C4), we can derive $\hatbeta_1,\renew_1$ is consistent estimator according to \citet{newey1987asymmetric}. Here, we prove the consistency of for an arbitrary $t \geq 2$.
	
	According to \citet{luo2020renewable}, we define $\score(\bbeta)$ as the score function of $\renewloss_{\Nb}(\bbeta)$:
	\begin{align*}
		\score(\bbeta)=\frac{\partial\renewloss_{\Nb}(\bbeta)}{\partial\bbeta}=\frac{1}{\Nb}\left\{\sumb W_{\nt}(\renew_t)(\bbeta-\renew_{b-1})+\nb \cdot\grad(\bbeta)\right\},
	\end{align*}
	where $\grad(\bbeta)$ is the gradient function of $\loss_{\nb}(\bbeta)$, defined as:
	\begin{align*}
		\grad(\bbeta)=\frac{\partial \loss_{\nb}(\bbeta)}{\partial \bbeta}=-\frac{1}{\nb}\sum_{i=1}^{\nb}|\tau-1(y_{bi}<\x_{bi}^\top \bbeta)|\cdot\x_{bi}(y_{bi}-\x_{bi}^\top \bbeta)=-\frac{1}{\nb}\Big[U_{\nb}(\bbeta)-W_{\nb}(\bbeta)\bbeta\Big].
	\end{align*}
	
	Note that when $\renew_{b-1}$ is consistent, we have
	\begin{align}\label{eq:score}
		\score({\truebeta})=\frac{1}{\Nb}\left\{\sumb W_{\nt}(\renew_t)(\truebeta-\renew_{b-1})+\nb\cdot\grad(\truebeta)\right\}=\op(1).
	\end{align}
	According to (\ref{eq:online-taylor}), the renewable estimator $\renew_b$ satisfies
	$\score(\renew_b)=\textbf{0}.$ Then we have
	\begin{align}\label{eq:score-diff}
		\score(\truebeta)=\score(\truebeta)-\score(\renew_b)= \frac{1}{\Nb}\left\{\sumb W_{\nt}(\renew_t)(\truebeta-\renew_{b})+\nb \cdot\left(\grad(\truebeta)-\grad(\renew_b)\right)\right\}=\op(1).
	\end{align}
	Taking the first-order Taylor series expansion of $\grad(\renew_b)$ around $\truebeta$ yields
	\begin{align}\label{eq:g}
		\grad(\renew_b)&=\grad(\truebeta)+\frac{1}{n_b}W_{\nb}(\barbeta_b)(\renew_b-\truebeta)\nonumber\\
		&=\grad(\truebeta)+\frac{1}{n_b}\left[W_{\nb}(\truebeta)+W_{\nb}(\barbeta_b)-W_{\nb}(\truebeta)\right](\renew_b-\truebeta),
	\end{align}
	where $\barbeta_b$ lies between $\truebeta$ and $\renew_b$. According to Lemma \ref{L1}, we know that
	\begin{align*}
		\frac{1}{n_b}\left\|W_{\nb}(\barbeta_b)-W_{\nb}(\truebeta)\right\|&=O_p(\|\barbeta_b-\truebeta\|)=O_p(\|\renew_b-\truebeta\|).
	\end{align*}
	Thus, rewrite (\ref{eq:g}) as
	\begin{align}\label{eq:g-1}
		\grad(\renew_b)&=\grad(\truebeta)+\frac{1}{n_b}W_{\nb}(\truebeta)(\renew_b-\truebeta)+O_p(\|\renew_b-\truebeta\|^2).
	\end{align}
	Combining (\ref{eq:score-diff}) and (\ref{eq:g-1}) yields
	\begin{align}\label{eq:score-diff2}
		\begin{aligned}
			\score(\truebeta)-\score(\renew_b)&= \frac{1}{\Nb}\left\{\sumb W_{\nt}(\renew_t)+W_{\nb}(\truebeta)\right\}(\truebeta-\renew_{b})\\
			& +\Op\left(\frac{\nb}{\Nb}\left\|\renew_{b}-\truebeta\right\|^2\right)=\op(1).
		\end{aligned}
	\end{align}
	As $\Nb \rightarrow \infty$, with (C3), we can know that $\Nb^{-1}\left\{\sumb W_{\nt}(\renew_t)+W_{\nb}(\truebeta)\right\}$ is positive definite. It follows that $\renew_b \xrightarrow{p} \truebeta$. 
	
	\subsection{Technical proofs of Theorem 2 (Asymptotic Normality)}\label{proof-sub3}
	When $N_1=n_1 \rightarrow \infty$ it is easy to derive that $\hatbeta_1$ equals to $\renew_1$ and satisfied $\sqrt{N_1}(\renew_1-\truebeta)=\sqrt{N_1}(\hatbeta_1-\truebeta) \xrightarrow{d}N\left(\textbf{0},\Sigmaw^{-1}\boldsymbol{\Omega}\Sigmaw^{-1}\right)$. In addition, the score function of first data batch has the following stochastic expression
	\begin{align}\label{eq:stochastic1}
		\boldsymbol{g}_{1}(\truebeta)&= -\frac{1}{n_1} W_{n_1}(\hatbeta_1)(\hatbeta_1-\truebeta)+\Op(\|\hatbeta_1-\truebeta\|^2).
	\end{align} 
	From eq~\eqref{eq:score} and~\eqref{eq:score-diff2}, we have
	\begin{align}\label{eq:score-diff3}
		\begin{aligned} \frac{1}{\Nb}\left\{\sumb W_{\nt}(\renew_t)+W_{\nb}(\truebeta)\right\}(\truebeta-\renew_{b})&-\frac{1}{\Nb}\sumb W_{\nt}(\renew_t)(\truebeta-\renew_{b-1})\\&-\frac{\nb}{\Nb}\cdot\boldsymbol{g}_b(\truebeta)+\Op\left(\frac{\nb}{\Nb}\|\renew_{b}-\truebeta\|^2\right) = \boldsymbol{0}.
		\end{aligned}
	\end{align}
	Same as~\eqref{eq:stochastic1} for renewable estimator $\renew_{b-1}$, we have
	\begin{align}\label{eq:stochastict}
		\frac{1}{N_{b-1}}\sumb n_t\boldsymbol{g}_{t}(\truebeta)&= -\frac{1}{N_{b-1}}\sumb \nt \cdot \Hess_{\nt}(\renew_t)(\renew_{b-1}-\truebeta)+\Op\left(\sumb \frac{n_t}{N_{b-1}}\|\renew_t-\truebeta\|^2\right)\nonumber\\
		&=-\frac{1}{N_{b-1}}\sumb W_{n_t}(\renew_t)(\renew_{b-1}-\truebeta)+\Op\left(\sumb \frac{n_t}{N_{b-1}}\|\renew_t-\truebeta\|^2\right).
	\end{align} 
	By replacing the second term in~\eqref{eq:score-diff3} with~\eqref{eq:stochastict}, we have
	\begin{align*}
		\frac{1}{\Nb}\sum_{t=1}^b n_t \cdot\boldsymbol{g}_t(\truebeta)-\frac{1}{\Nb}\left\{\sumb W_{\nt}(\renew_t)+W_{\nb}(\truebeta)\right\}(\truebeta-\renew_b)+\Op\left(\sum_{t=1}^b\frac{n_t}{\Nb}\|\renew_t-\truebeta\|^2\right)=\boldsymbol{0}.
	\end{align*}
	With Theorem~\ref{T1}, we have $\renew_t\xrightarrow{p}\truebeta$ for $t=1,2,\cdots,b-1$, and by (C3), it holds
	\begin{align}\label{eq:allscore}
		\frac{1}{\Nb}\sum_{t=1}^b n_t\cdot\boldsymbol{g}_t(\truebeta)+\frac{1}{\Nb}\sum_{t=1}^b W_{\nt}(\truebeta)(\renew_b-\truebeta)+\Op\left(\sum_{t=1}^b\frac{n_t}{\Nb}\|\renew_t-\truebeta\|^2\right)=\boldsymbol{0},
	\end{align}
	which implies that 
	\begin{align*}
		\sqrt{\Nb}(\renew_b-\truebeta)=\Bigg[\frac{1}{\Nb}\sum_{t=1}^b W_{\nt}(\truebeta)\Bigg]^{-1}\frac{-1}{\sqrt{\Nb}}\sum_{t=1}^b \nt\cdot\boldsymbol{g}_t(\truebeta)+\op(1)
	\end{align*}
	With weak law of large number, we have $\frac{1}{\Nb}\sum_{t=1}^b W_{\nt}(\truebeta) \xrightarrow{p}\Sigmaw$, and according to central limit theorem, when $\Nb \rightarrow \infty$ we have
	\begin{align*}
		\frac{-1}{\sqrt{\Nb}}\sum_{t=1}^b\nt\cdot\boldsymbol{g}_t(\truebeta)=\frac{1}{\sqrt{\Nb}}\sum_{t=1}^b\sum_{i=1}^{\nt}|\tau-1(y_{ti}<\x_{ti}^\top\truebeta)|\cdot\x_{ti}(y_{ti}-\x_{ti}^\top\truebeta)\xrightarrow{d}N\left(\textbf{0},\boldsymbol{\Omega}\right).
	\end{align*}
	Hence, the renewable estimator $\renew_b$ satisfies the following asymptotic normal distribution
	$$\sqrt{\Nb}\big(\renew_b-\truebeta \big)\xrightarrow{d} N\left(\textbf{0},\Sigmaw^{-1}\boldsymbol{\Omega}\Sigmaw^{-1}\right).$$
	
	\subsection{Technical proofs of Theorem 3 (Asymptotic equivalency)}\label{proof-sub4}
	Now we prove the asymptotic equivalency of $\renew_b$ and $\oraclebeta$. According to Theorem~\ref{T1}, we know that all $\renew_t$ are consistent for $t = 1,2,\cdots,b$. Hence, it is easy to derive that $W_{\nb}(\renew_{b-1})\xrightarrow{p}W_{\nb}(\renew_{b}),U_b(\renew_{b-1})\xrightarrow{p}U_b(\renew_{b})$ which implies that the approximate estimator define in~\eqref{eq:renewable} is equivalent to the estimator $\renew_b$ derive from~\eqref{eq:online-sol}.
	Due to $\oraclebeta$ is the static estimator satisfied $\sum_{t=1}^b \nt \cdot \boldsymbol{g}_{t}(\oraclebeta)=\boldsymbol{0}$, by taking the first-order Taylor series expansion around $\truebeta$ we have
	\begin{align} \label{eq:score-oracle}
		\frac{1}{\Nb} \sum_{t=1}^b \nt \cdot \boldsymbol{g}_{t}(\oraclebeta)= \frac{1}{\Nb} \sum_{t=1}^b \nt \cdot \boldsymbol{g}_{t}(\truebeta)+\frac{1}{\Nb}\sum_{t=1}^bW_{\nt}(\truebeta)(\oraclebeta-\truebeta)+\Op(\|\oraclebeta-\truebeta\|^2)=\boldsymbol{0}.
	\end{align}
	In addition, take difference between \eqref{eq:allscore} and~\eqref{eq:score-oracle}, it holds
	\begin{align*}
		\frac{1}{\Nb}\sum_{t=1}^bW_{\nt}(\truebeta)(\renew_b-\oraclebeta)=\Op\left(\sum_{t=1}^b\frac{\nt}{\Nb}\left\|\renew_t-\truebeta\right\|^2+\|\oraclebeta-\truebeta\|^2\right)=\Op\left(\frac{1}{\Nb}\right).
	\end{align*}
	Under Theorem~\ref{T2}, we have $\|\renew_b-\truebeta\|^2=\Op(1/N_b)$, hence it holds
	$$\|\renew_b-\oraclebeta\| = O_p(1/\Nb).$$
	
	\section{Additional Simulations}\label{app:sim}
\begin{figure}
		\centering
		\includegraphics[width = \linewidth]{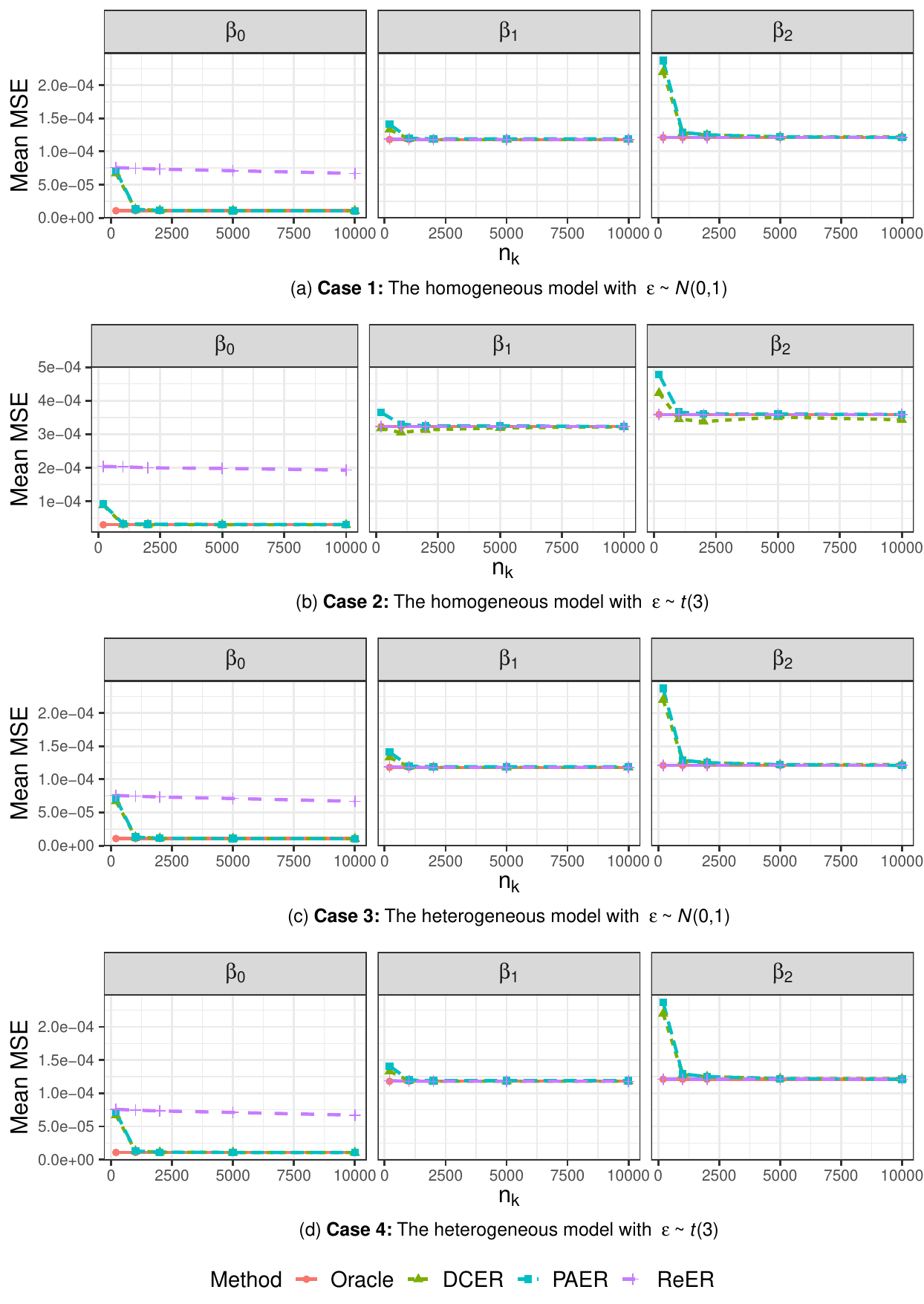}
		\caption{MSE values for fixed $N_k$ with varying $n_k$ at the $50\%$ expectile level.}
		\label{fig:fix-50}
	\end{figure}

\begin{figure}
		\centering
		\includegraphics[width = \linewidth]{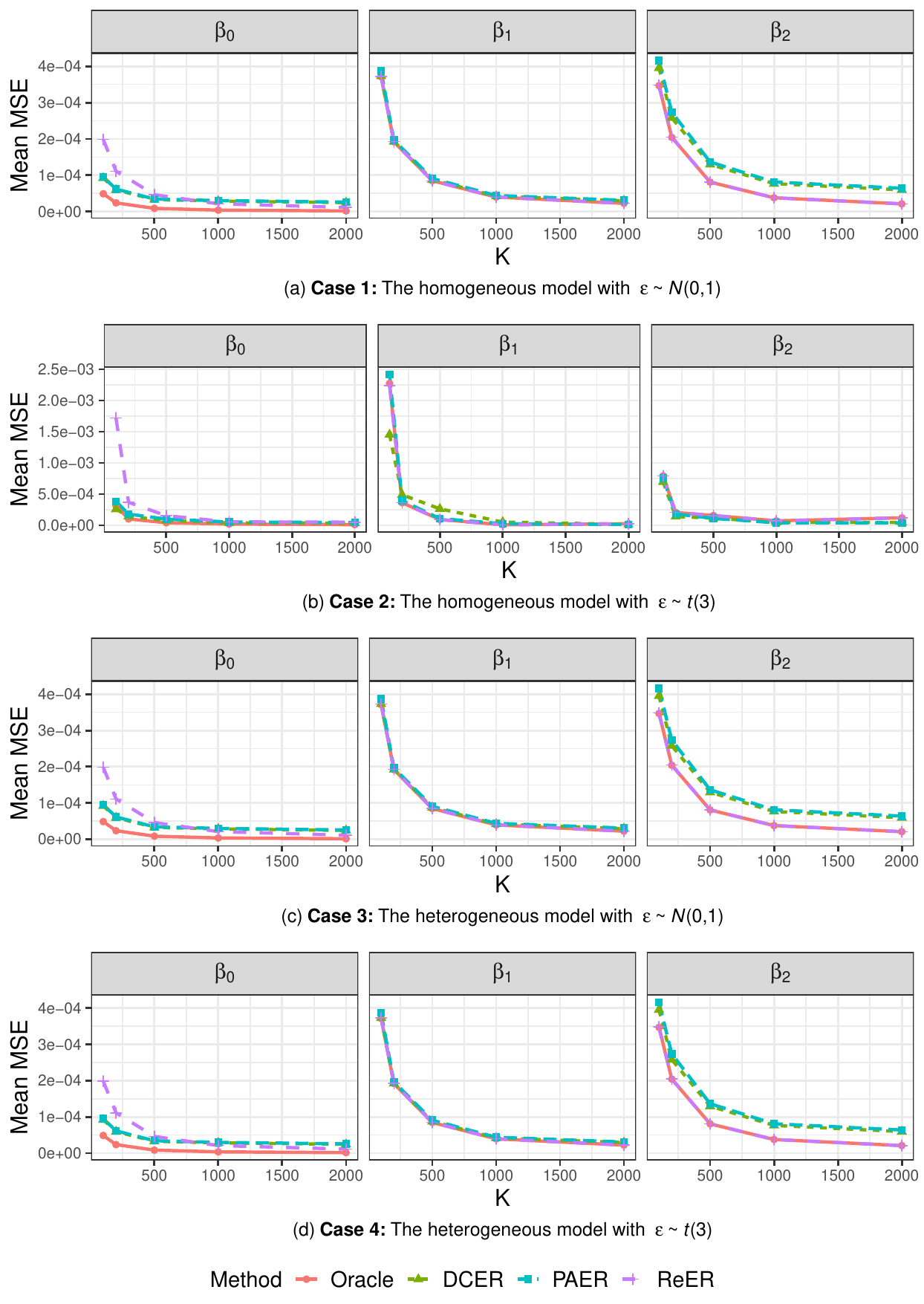}
		\caption{MSE values for fixed $n_k$ with varying $K$ at the $50\%$ expectile level.}
		\label{fig:stream-50}
	\end{figure}

\begin{figure}
		\centering
		\includegraphics[width = \linewidth]{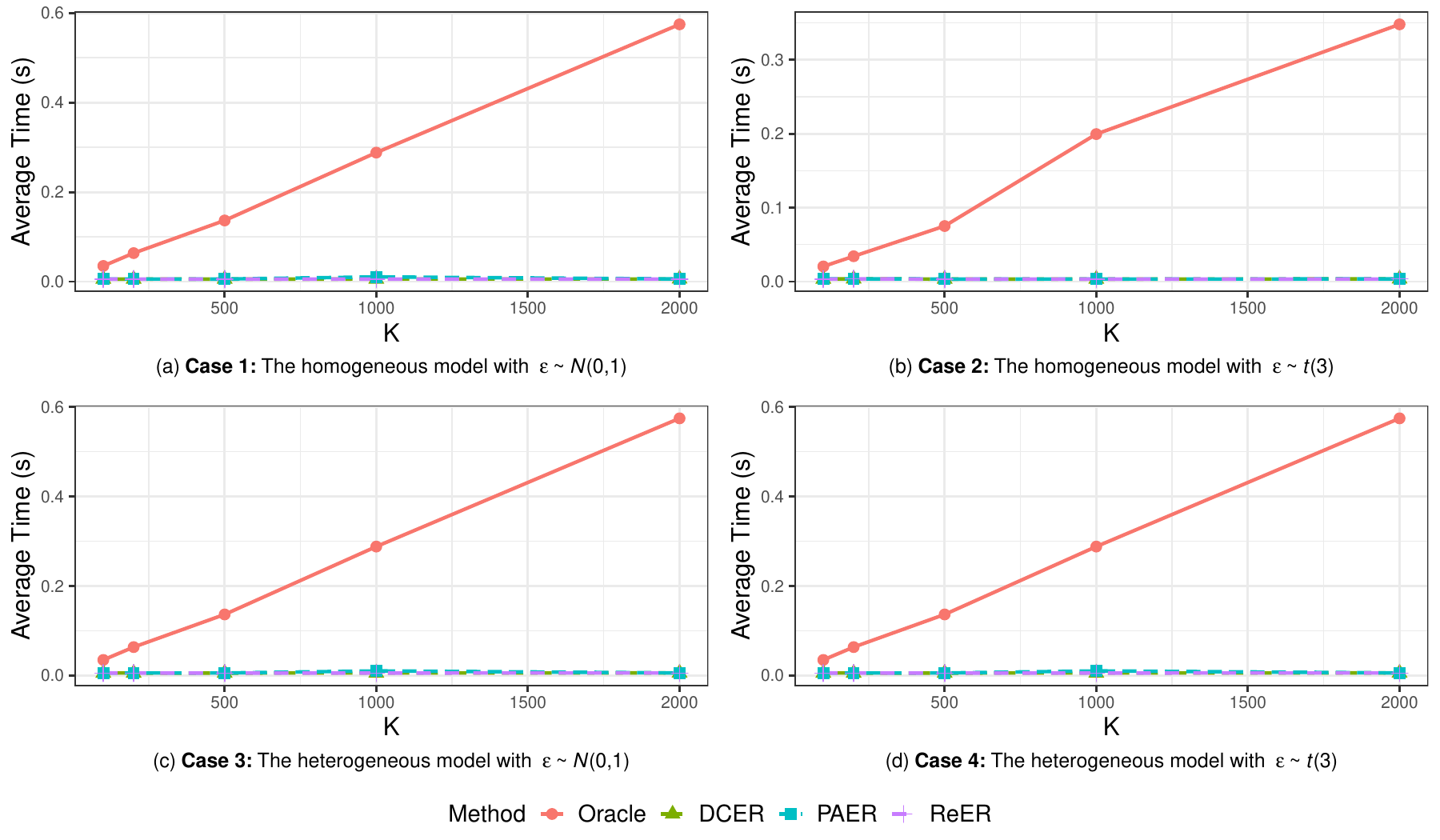}
		\caption{The computational time of renewable method and the Oracle method with fixed $n_k$ and varying $K$ in 50\% expectile level.}
		\label{fig:time-50}
	\end{figure}

\begin{figure}
		\centering
		\includegraphics[width = \linewidth]{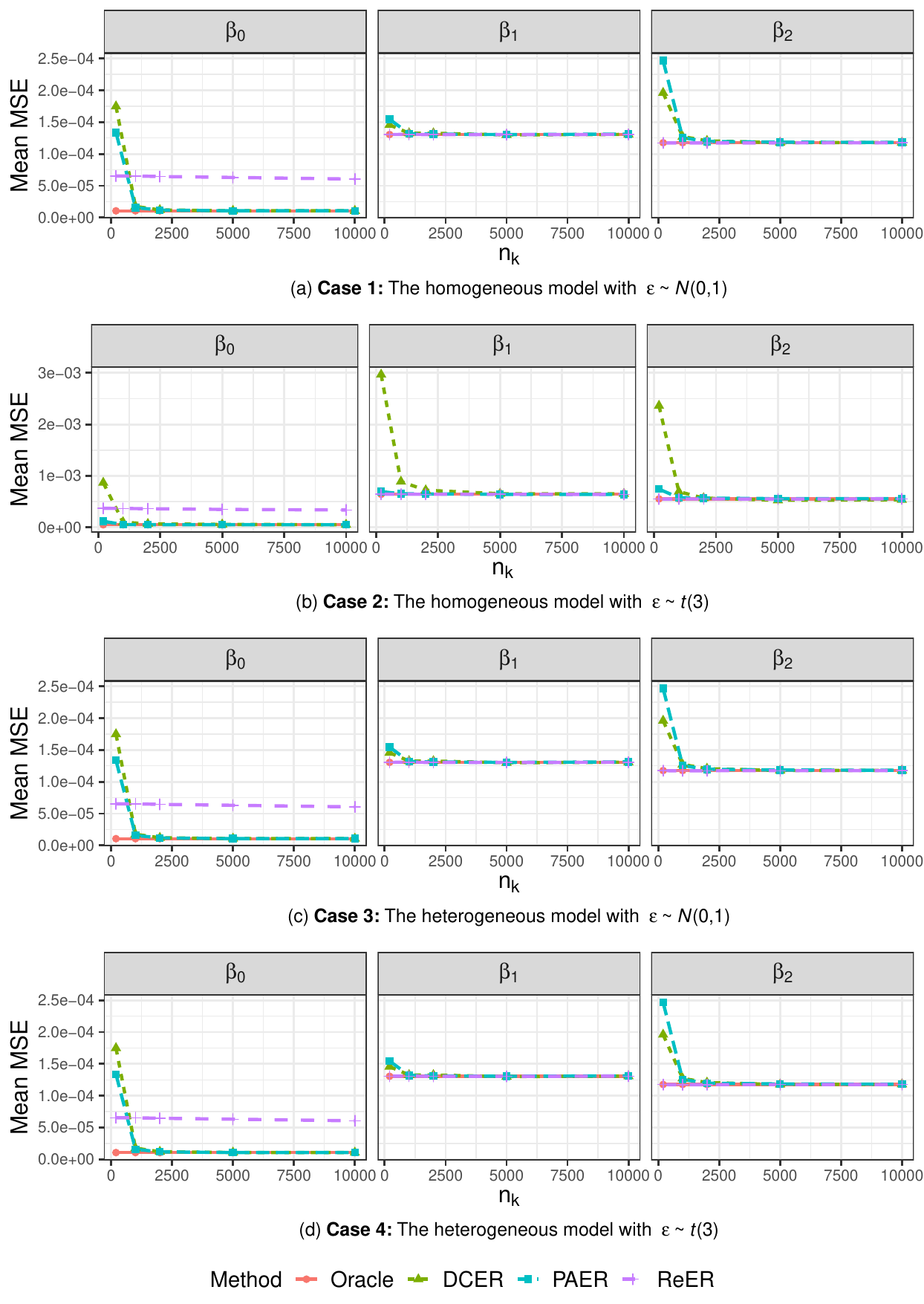}
		\caption{MSE values for fixed $N_k$ with varying $n_k$ at the $75\%$ expectile level.}
		\label{fig:fix-75}
	\end{figure}

\begin{figure}
		\centering
		\includegraphics[width = \linewidth]{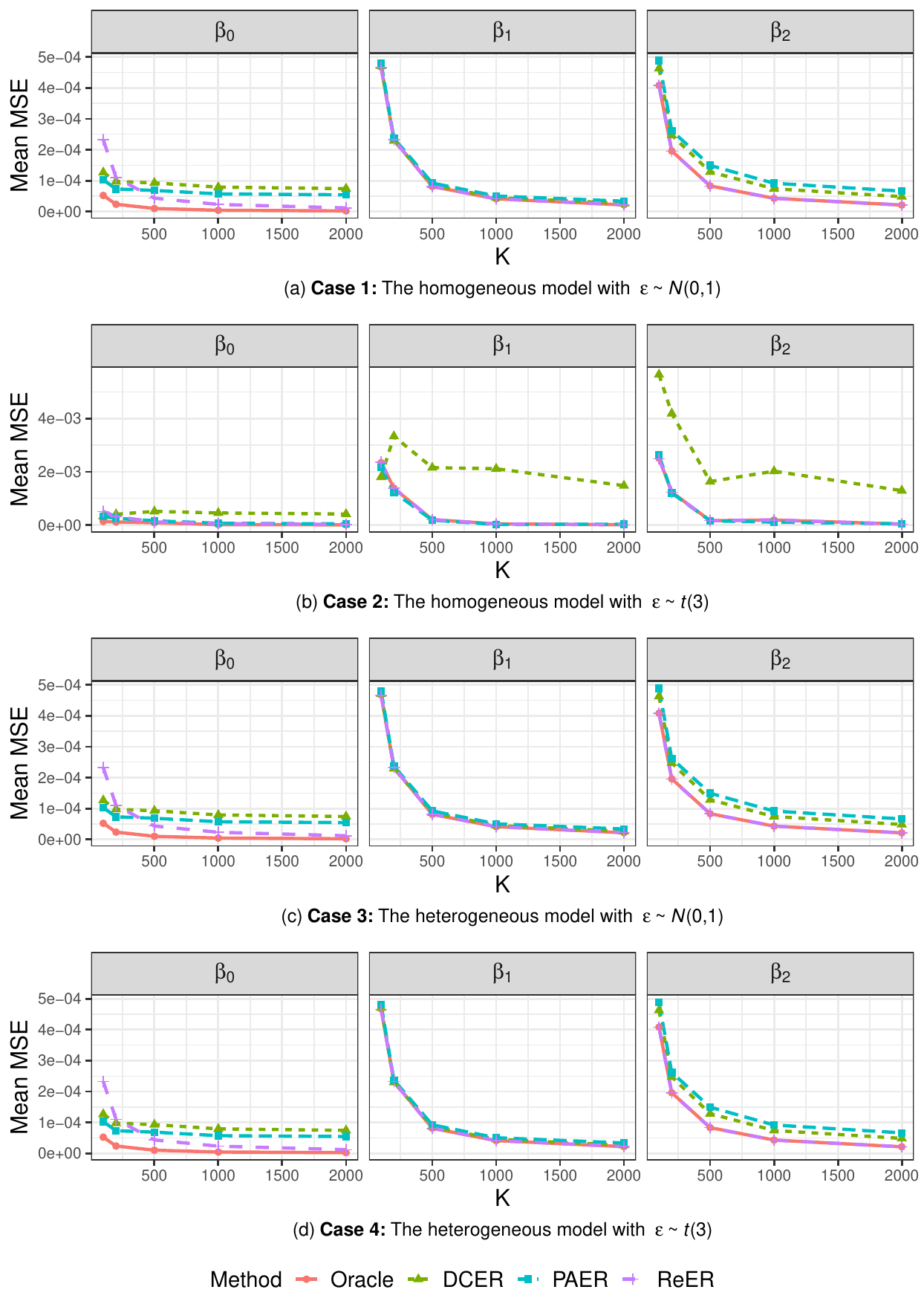}
		\caption{MSE values for fixed $n_k$ with varying $K$ at the $75\%$ expectile level.}
		\label{fig:stream-75}
	\end{figure}

\begin{figure}
		\centering
		\includegraphics[width = \linewidth]{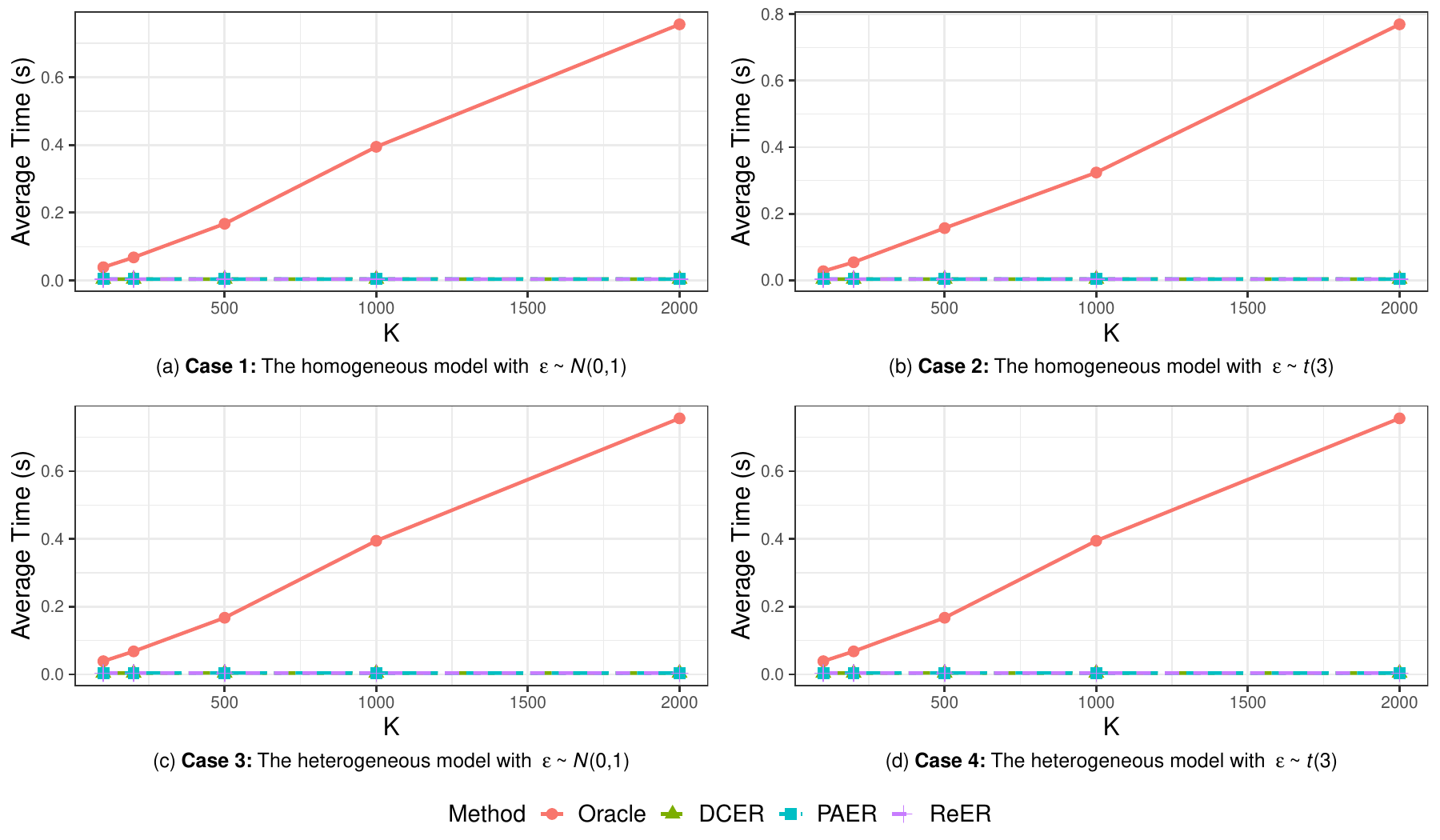}
		\caption{The computational time of renewable method and the Oracle method with fixed $n_k$ and varying $K$ in 75\% expectile level.}
		\label{fig:time-75}
	\end{figure}

	The performance results at the 50\% and 75\% expectile levels are summarized below. Figures~\ref{fig:fix-50} and~\ref{fig:stream-50} illustrate the MSEs under Scenario \textbf{S1} (fixed $N_k$ with varying $n_k$) and Scenario \textbf{S2} (fixed $n_k$ with varying $K$), respectively, at the 50\% expectile level. Likewise, Figures~\ref{fig:fix-75} and~\ref{fig:stream-75} present the corresponding results for the 75\% expectile level. Figures~\ref{fig:time-50} and~\ref{fig:time-75} show the computational time for the three renewable methods and the Oracle method at the 50\% and 75\% expectile levels, respectively.
	
\end{document}